\documentstyle[11pt,aaspp4]{article}
\tighten
\journalid{}{}
\articleid{}{}
\def\lesssim{\mathrel{\hbox{\rlap{\hbox{\lower4pt\hbox{$\sim$}}}\hbox{$<$}}}}
\def\gtrsim{\mathrel{\hbox{\rlap{\hbox{\lower4pt\hbox{$\sim$}}}\hbox{$>$}}}}
\def\sq{\hbox{\rlap{$\sqcap$}$\sqcup$}}
\newcommand{\ropt}{R$_{25}$ }
\newcommand{\ha}{H$\alpha$ }
\begin{document}
\title{THE EXTREME OUTER REGIONS OF DISK GALAXIES: I. CHEMICAL ABUNDANCES OF 
HII REGIONS}
\author{Annette~M.~N. Ferguson\altaffilmark{1,2,3}}
\affil{Department of Physics and Astronomy, The Johns Hopkins University,
    Baltimore, MD 21218}
\author{J. S. Gallagher\altaffilmark{2}}
\affil{Department of Astronomy, University of Wisconsin, 
Madison, WI 53706}
\author{Rosemary~F.~G. Wyse\altaffilmark{2,3}}
\affil{Department of Physics and Astronomy, The Johns Hopkins University,
    Baltimore, MD 21218}
\altaffiltext{1}{ Current Address: Institute of Astronomy, University of
Cambridge, Madingley Road, Cambridge, UK CB3 0HA}
\altaffiltext{2}{Visiting Astronomer, Kitt Peak National Observatory. 
KPNO is operated by AURA, Inc.\ under contract to the National Science
Foundation.}
\altaffiltext{3}{Visiting Astronomer, Lowell Observatory.}

\begin{abstract}
We present the first results of an ongoing project to investigate the
present-day chemical abundances of the extreme outer parts of galactic
disks, as probed by the emission line spectra of a  new sample of HII
regions.   The galaxies studied here, NGC~628, NGC~1058 and NGC~6946,
are all late-type spiral galaxies, characterized by larger than average
HI-to-optical sizes. Our  deep H$\alpha$ images have revealed the
existence of recent massive star formation, traced by HII regions, out
to, and beyond, two optical radii in these galaxies (defined by the
B-band 25th magnitude isophote).  Optical spectra of these
newly-discovered HII regions are used to investigate their densities,
ionization parameters, extinctions and in particular their oxygen and
nitrogen abundances.  Our measurements reveal gas-phase abundances of
O/H~$\sim$~10-15\%~ of the solar value, and N/O~$\sim$~20-25\%~ of the
solar value, at radii of 1.5--2~R$_{25}$.  Clear evidence also exists
for diminished dust extinction (A$_V$~$\sim$~0--0.2) at large radii.

The combination of our measurements of outer disk HII region abundances
with those for inner disk HII regions published in the literature is a
powerful probe of the shape of abundance gradients over unprecedented
radial baselines.  The predictions of models of chemical evolution
often diverge most strongly in the outer parts of galaxies.  Both the
oxygen and the nitrogen-to-oxygen abundances generally decrease with
increasing radius. Within the limits of the current dataset, the
radial abundance variations are consistent with  single log-linear
relationships, although the derived slopes can often differ
considerably from those found if only inner disk HII regions are used
to define the fit.  The small number of HII regions in our present
sample, together with uncertainties in the calibrations of the
empirical methods used here to determine abundances,  limit the ability
to constrain both subtle changes in the radial gradient and intrinsic
scatter at a fixed radius.  Nitrogen-to-oxygen ratios appear to be
consistent with a combination of primary and secondary production of
nitrogen.  Interestingly, both the mean level of enrichment and the
ratio of N/O measured in extreme outer galactic disks are similar to
those values measured in some high redshift damped Lyman-$\alpha$
absorbers, suggesting that outer disks at the present epoch are
relatively unevolved.
\end{abstract}


\section{Introduction}

Radial variations in the abundances of elements within galaxies are
well-established as providing important constraints on models of disk
galaxy formation and  evolution (eg. Pagel \& Edmunds 1981; Vila-Costas
\& Edmunds 1992; Zaritsky, Kennicutt \& Huchra 1994 (hereafter ZKH);
Prantzos \& Aubert 1995).  HII regions play a unique role in such
studies because they yield relatively reliable elemental abundances
from measurements of emission line intensities.  The early work of
Searle (1971) and Shields (1974) used observations of HII regions in
spirals to establish the existence of negative radial gradients  in the
abundance of oxygen.  Subsequent work has shown that such gradients are
a generic feature of disk galaxies, although the magnitude and the
shape of the gradient, as well the characteristic abundance,  are
observed to vary considerably from galaxy to galaxy (eg. Vila-Costas \&
Edmunds 1992; ZKH).

Unfortunately, most HII region abundance studies carried out to date
have probed only the bright, easily-observed inner regions of galactic
disks, lying at or within the classical optical radius, R$_{25}$
(defined by the B-band 25th magnitude isophote).  It is well known,
however,  that disk galaxies have HI  disks which extend to typically
$\gtrsim$ 1.5--2~\ropt (eg. Cayette et al 1994; Broeils 1994), and in
some rare cases to $\gtrsim$~3\ropt (eg. van der Kruit \& Shostak
1984).   These outer regions are characterized by low HI columns, high
gas fractions and long dynamical timescales and thus provide an
opportunity to study star formation and chemical evolution in rather
unique physical environments.  Indeed, the outer regions of disks have
physical properties which are reminiscent of those thought to exist
during the early stages of galaxy formation, as well as those in giant
low surface brightness galaxies, such as Malin 1 (eg. Pickering et al
1997).  Studying  the mean enrichment level, the shape of the abundance
gradient and the amount of intrinsic scatter at fixed radius, in outer
galactic disks  will therefore forward our understanding of a broad
range of astrophysical objects.  Furthermore, since the predictions of
models of chemical evolution often diverge most strongly in the outer
parts of galaxies, abundance determinations extending as far in the
outer disk as possible are needed to discriminate between competing
theories.

Only a handful of outer disk HII regions have measured abundances and,
of these, very few lie at galactocentric radii significantly beyond the
optical radius (eg.  Garnett \& Shields 1987; Garnett et al 1992;
Garnett \& Kennicutt 1994).  The bias of previous studies of HII
regions to the inner disk  has certainly been due in part to the lack
of known HII regions lying at large radii.  In the course of a large,
deep, H$\alpha$ imaging survey to search for and study star formation
in the extreme outer regions of disk galaxies (Ferguson 1997), we have
discovered numerous faint HII regions in the low gas surface density
(N$_{HI}~\lesssim$~few~$\times~10^{20}$~cm$^{-2}$) outer limits  of
several galaxies.  These HII regions are typically small
(diameter~$\sim$~150--500~pc) and of low luminosity
(L$_{H\alpha}$~$\sim$~1--10~L$_{Orion}$), and do not clump into the
giant complexes which populate the inner disks  of spiral galaxies.
They are often observed to trace out narrow spiral arms, which are
sometimes coincident with underlying HI arms and faint stellar arms
($\mu_B \sim$ 26--28 mag/$\sq\arcsec$).  A study of the ongoing and
past  star formation in the outer disks of these galaxies and its
implications for models of disk galaxy evolution will be presented
elsewhere (Ferguson et al, 1998a,b).

We present here the first results from a long-slit spectroscopic study
to determine the chemical abundances in the extreme outer parts of nearby
disk galaxies, using the emission line spectra of our newly discovered
HII regions.  This paper presents results for three late-type spiral
galaxies, namely NGC~628, NGC~1058 and NGC~6946.   These galaxies have
particularly extended disks of neutral hydrogen and are amongst our
best examples of galaxies with extreme outer disk star formation.   A
summary of their properties is presented in Table 1.  Chemical
abundances for HII regions in the inner disks of NGC~628 and NGC~6946
have been derived by McCall, Rybski \& Shields (1985: hereafter MRS).
Chemical abundances have not been previously measured for HII regions
in NGC~1058, and we present here a study of both inner and outer disk
HII regions in that galaxy.

\section{Observations}

The typical H$\alpha$ fluxes of the outer HII regions  we have
identified range from 1--70$\times10^{-15}$~erg~s$^{-1}$~cm$^{-2}$,
which for reference is roughly 10--1000 times fainter than the usual
HII regions in late type spirals which have  previously been studied
spectroscopically (eg.  MRS, ZKH).  Our strategy was to first target
the  brightest outer disk HII regions in our sample.  The data
discussed here  were obtained during November 1994, using the KPNO 4m
telescope and the RC-spectrograph + T2KB.  The T2KB chip was run with a
gain of 2e$^{-}$/ADU and readnoise of $\sim$ 4e$^{-}$.   We used a
2{\arcsec} x 300{\arcsec} slit with a single moderate-resolution
grating (KPC10-A; 316 l/mm) to cover the wavelength range of range
3600-7500{\AA}, with spectral resolution of
$\sim$~7{\AA}.\footnote{During a previous observing run, the data from
which will be reported in a future paper, we  used higher spectral
resolution, requiring two gratings,  in an attempt to  detect the
temperature-sensitive [OIII] $\lambda$4363 line. Such resolution was
required in order to seperate the [OIII] $\lambda$4363 line from any
scattered HgI $\lambda$4358 emission.  Unfortunately, we failed to
detect [OIII] $\lambda$4363 at all and so switched strategies to a
simpler observing mode.}  The  grating was used in first order with an
order-blocking filter (WG345) to eliminate second-order blue light from
below $\sim$ 3200{\AA}.  The seeing was typically 1.5--2\arcsec.  Of
the three nights of our run, two nights were mostly clear, hampered
only by light to moderate  cirrus during the latter parts of each
night, but the third night was completely lost.  In this paper, we
discuss only a subset of the data obtained during the November run,
consisting of measurements of HII regions in the disks of NGC~628,
NGC~1058 and NGC~6946.

Blind offsets were required for all of our target outer HII regions.
These were derived from deep narrow-band images obtained using the KPNO
0.9m and Lowell 1.8m telescopes, and were based on either nearby bright
stars or the nucleus of the galaxy.  The images were astrometrically
calibrated using a grid of stars measured on the digitised POSS plates
and are accurate to $\lesssim$~0.5\arcsec.  Individual exposure times
were 30 min per region and we usually obtained 2--4 exposures per
target region, depending on the faintness of the target.  Experience
indicated that if no metal lines were detected in a single 30 min
exposure, then there was negligible chance of obtaining useful data for
abundance determinations by stacking additional exposures.  Most
objects were observed at low airmass and we rotated the slit to
parallactic angle whenever possible to eliminate the loss of blue
light.   We also attempted to position the slit in such a manner as to
include as many HII regions as possible, but the small angular size,
faintness and relative isolation of our outermost targets made this
difficult.  Spectra of He-Ne-Ar lamps were otained before and after
each object exposure, and we made multiple observations of standard
stars from the list of Massey et al (1988).

In Figures \ref{628ima}--\ref{6946ima}, we show continuum--subtracted
H$\alpha$ images of each galaxy with the target HII regions
identified.  Table 2 lists the identifications, positions and
properties of our sample of HII regions.  The diameters and fluxes
presented are approximate  and are intended simply to illustrate the
range in physical properties which exists within our sample.  In
particular, it is very difficult (and subjective) to assign sizes and
fluxes to inner disk HII regions, which often are part of large
complexes composed of many $`$cores'.  We have not corrected the fluxes
and luminosities for the effects of internal extinction.  Most of the
outer disk HII regions in our sample have luminosities consistent with
their   being ionized by only a few equivalent O5V stars, as calculated
using the Lyman continuum fluxes in Vacca et al (1996); these estimates
are strictly lower limits to the number of enclosed massive stars as
the HII regions  may well  not be ionization bounded (eg.  see the
discussion of the origins of diffuse ionized gas by Ferguson et al
1996a,b). Two inner disk HII regions from the sample of MRS are
included in our sample in order to provide  an external consistency
check on our measurements.

\newpage

\centerline{see 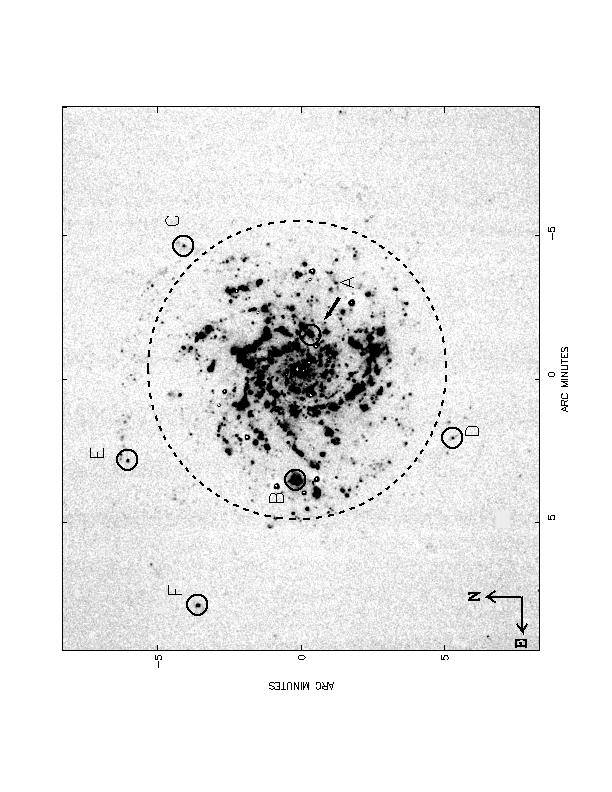}
\figcaption{An \ha continuum--subtracted image of NGC~628 obtained
using the KPNO 0.9m, with the target HII regions identified.  R$_{25}$
is marked.  North is to the top and east to the left.  \label{628ima} }
\bigskip
\bigskip
\centerline{see 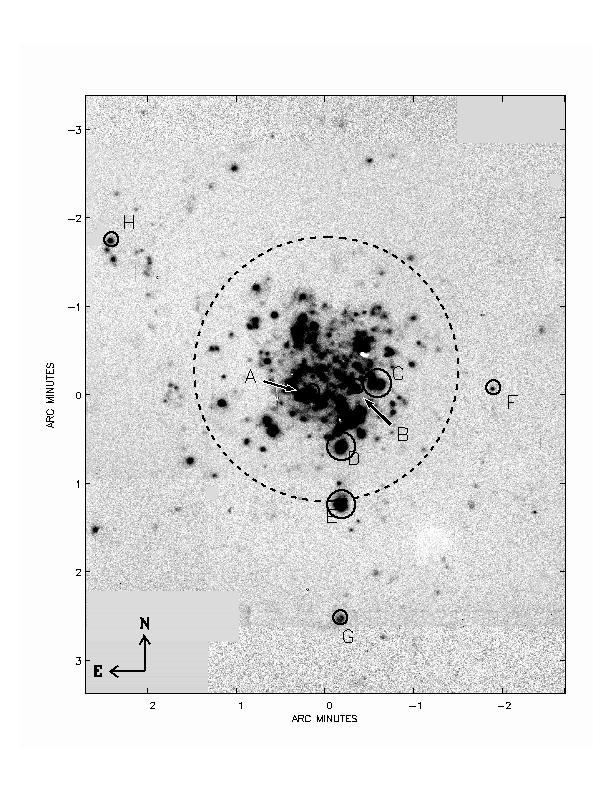}
\figcaption{A mosaiced \ha continuum--subtracted image of NGC~1058 obtained
using the Lowell 1.8m, with the target HII regions identified.  R$_{25}$
is marked.  North is to the top and east to the left.  \label{1058ima} }
\bigskip
\bigskip
\centerline{see 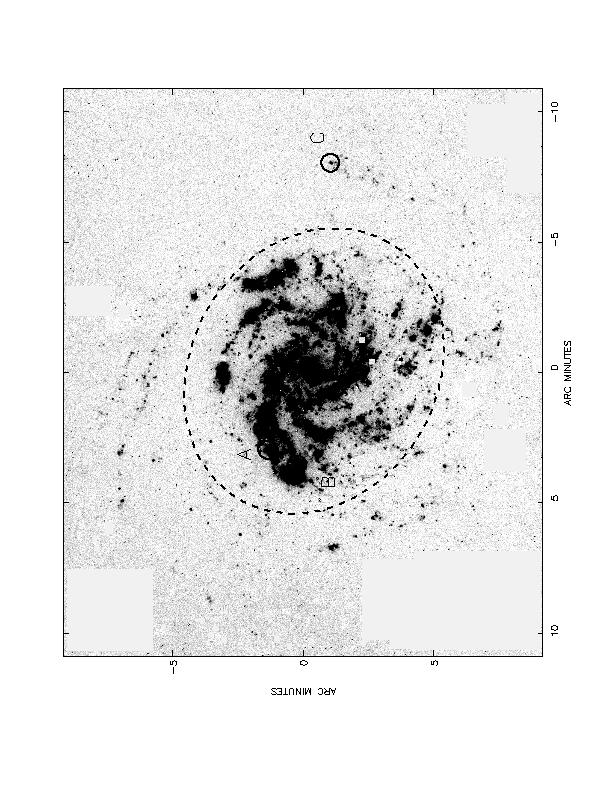}
\figcaption{An \ha continuum--subtracted image of NGC~6946 obtained
using the KPNO 0.9m, with the target HII regions identified.  R$_{25}$
is marked.  North is to the top and east to the left.  \label{6946ima} }

\newpage

\section{Data Analysis}  

Preliminary data reduction was carried out using standard
procedures.   A DC offset was subtracted from each frame using the
overscan region. Bias frames and quartz lamp exposures were used to
remove any residual structure  in the DC offset and pixel-to-pixel gain
variations respectively.  Dark frames were also obtained but the dark
current was found to be negligible and the frames were not utilized in
the analysis.   Exposures of the twilight sky were obtained to map out
the illumination pattern along the slit, and flatten the data in the
spatial direction.

\begin{figure}
\plotone{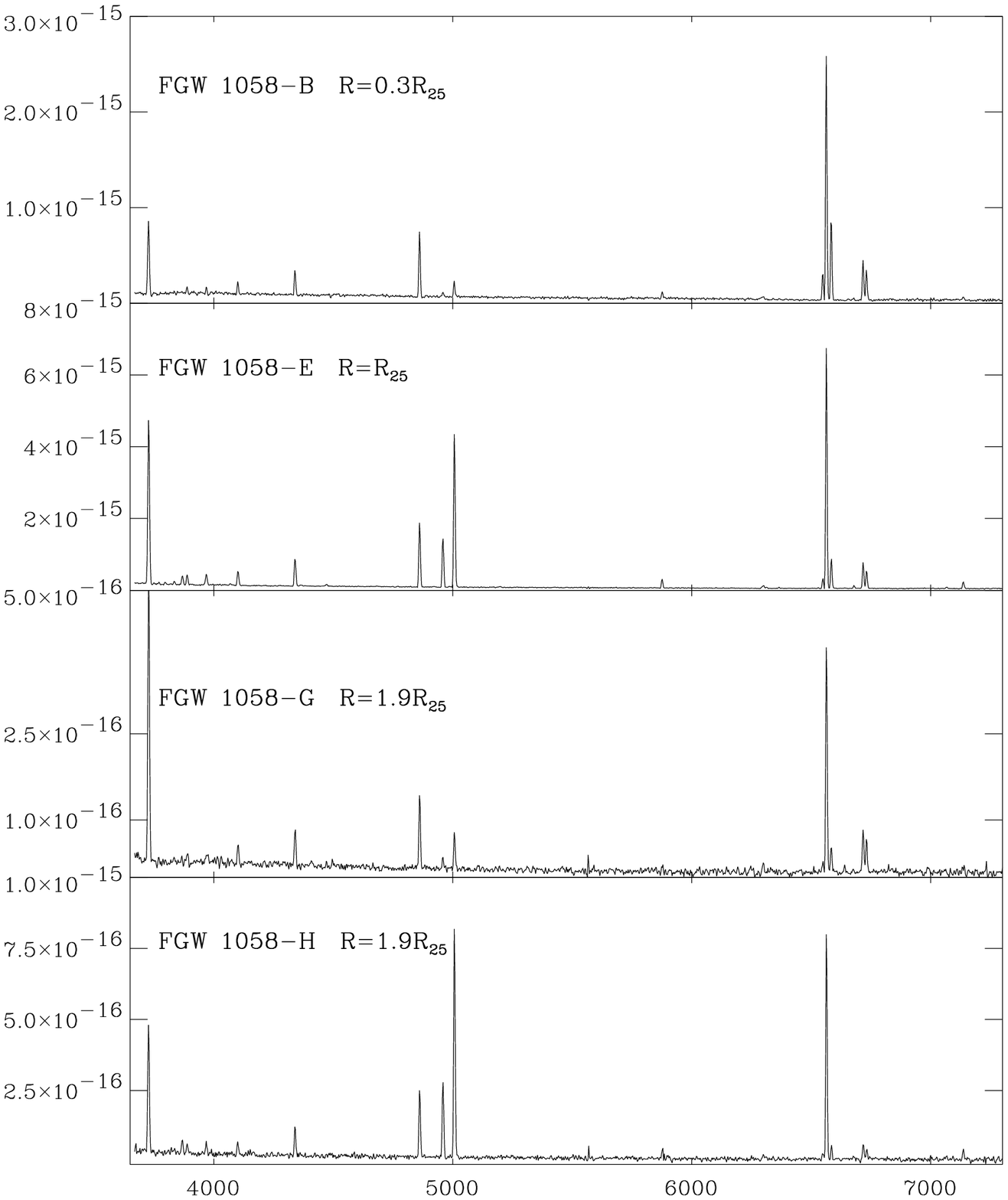}
\caption{Representative spectra from this study.  Shown are four of
the HII regions observed in NGC~1058, displayed in order of increasing
galactocentric radius (and metal abundance).  The last two spectra
represent HII regions at the same radius, which exhibit different
excitations.   The units are
erg~s$^{-1}$~cm$^{-2}$~{\AA}$^{-1}$.\label{refspect}}
\end{figure}

One dimensional spectra were extracted with apertures ranging in size
from 3--15\arcsec, depending on the seeing,  and on the size and
brightness of the object in question.  The sky to be subtracted was
selected from adjacent regions, and fit with a low order polynomial.
An important step in the extraction process is mapping the location of
the spectrum at each point along the dispersion axis.  In the brightest
HII regions observed, this $`$trace' could be well--defined, and
revealed only a small amount of distortion from the blue to the red,
which was typically much less than the size of the extraction box.
Only a weak continuum, if any, was present in the outer disk HII
regions however.  In those cases where we could not trace the continuum
along the entire dispersion axis, we chose to fit a low order
polynomial anchored to the position of one of the brightest lines
present.   The individual extracted spectra were wavelength calibrated
using He-Ne-Ar exposures, with the typical accuracy of the
transformation being a few tenths of a pixel, or equivalently $\sim$0.5
{\AA}.  Flux calibration was carried out using observations of standard
stars (with the same slit width as our program objects), and the
sensitivity functions showed residuals of  only $\sim$0.02--0.03
magnitudes (after a zeropoint offset).  At this point, individual
spectra of the same HII region were checked for consistency and then
averaged together.   Spectra that had been dispersion corrected, but
not flux calibrated, were also averaged for the purpose of computing
uncertainties in the line intensities.   In Figure \ref{refspect}, we
present extracted, combined, calibrated spectra for 4 HII regions in
NGC~1058, which are representative of the typical spectra obtained in
this study.

Inspection of the spectra reveals the presence of many emission
lines.   Most prominent are the bright oxygen, nitrogen and sulfur
lines, as well as the Balmer lines of hydrogen.  In addition, several
HII region spectra show detections of fainter lines, such as those due
to helium, argon, neon and in some cases neutral oxygen.   We focus
here on only those bright lines which lead to determinations of the
oxygen and nitrogen abundances  through the semi-empirical methods
discussed below in section 4.2.  Emission line fluxes were measured via
gaussian fits to the line profiles.  The logarithmic extinction at
H$\beta$, C(H$\beta$), was derived from measurements of the Balmer
lines, using the equation
$$\mathrm
\frac{I_{\lambda}}{I_{H\beta}}=\frac{F_{\lambda}}{F_{H\beta}}~10^{{C(H\beta)}{f(\lambda)}}$$
where I$_{\lambda}$ is the intrinsic line flux, F$_{\lambda}$ is the
observed line flux, and f($\lambda)$ is the Galactic reddening function
normalized to H$\beta$.  The reddening function of Seaton (1979), was
adopted, as parametrized by  Howarth (1983), and assuming
R=A$_{V}$/E(B$-$V)=3.1.  Intrinsic Balmer line ratios were taken from
Osterbrock (1989), assuming an electron density of N$_e$=100 cm$^{-3}$
and an electron temperature T$_e$=10$^4$~K.  The determination of the
amount of underlying stellar Balmer absorption is an important concern
in the estimation of the required extinction correction.  Our data are
of insufficient S/N to determine the absorption equivalent width
directly, so we adopted the standard approach of assuming a correction
of 2{\AA} to the measured equivalent width due to underlying stellar
absorption (eg.  MRS; Oey \& Kennicutt 1993),  and we proceeded to
derive the logarithmic extinction at H$\beta$  based on the H$\beta$
and H$\alpha$ lines alone.   We also corrected the
forbidden-line/H$\beta$ ratios for the effects of underlying H$\beta$
absorption, but this is generally a small effect since most spectra
have moderate to large H$\beta$ equivalent widths.

Formal  errors in the derived line ratios were determined by summing in
quadrature the statistical noise from the number of counts, the
uncertainty in the continuum placement (proportional to the width of
the line times the RMS in the nearby continuum, corrected for the
effects of pixelization), and the uncertainty in the flux
calibration.   In addition, the error in the C(H$\beta$) term  was
accounted for when deriving extinction-corrected line ratios.  Tables
3--5 present the observed line intensities, both uncorrected and
corrected for reddening and Balmer absorption, as well as some relevant
line-ratios for  our sample of HII regions.  Formal errors on the
quantities are indicated in parentheses.

\section{Deriving Nebular Abundances}  
\subsection{The Direct Method}

The $`$direct' method for determining chemical compositions from
nebular emission lines  requires a knowledge of the electron
temperature (and density) of the emitting gas in order to transform
reddening-corrected emission line ratios to ionic abundance ratios, and
finally to elemental abundances  (eg. Osterbrock 1989).  The electron
temperature T$_e$ is commonly derived from the O$^{++}$ ion,  via the
ratio
$$\mathrm\frac{\left[OIII\right] \lambda 4959+\left[OIII\right] \lambda 5007}{\left[OIII\right] \lambda 4363}=\frac{7.73\exp^{(3.29{\times}10^4)/T}}{1+4.5{\cdot}10^{-4}(N_e/T^{1/2})}$$
(Osterbrock 1989).  Unfortunately, the temperature sensitive [OIII] $\lambda$4363
line is typically very weak in extragalactic HII regions, and decreases
in strength rapidly with increasing abundance; as a result, this method
is limited to only the hottest (ie. most metal poor) and brightest
objects.  The faintness of our outer disk HII regions, coupled with the
low resolution of our spectroscopy, severely limits the detectability
of this key diagnostic line.   A marginal detection of [OIII]
$\lambda$4363 was made in one of the individual spectra  we obtained
for the brightest outer disk HII region of our sample, 1058--E, but the
significance of the detection is low.  It can be used only to place an
upper limit on the electron temperature of the region, and hence a
lower limit on the oxygen abundance.  We measure [OIII]
$\lambda$4363/H$\gamma$~$\lesssim$~0.067, which, when combined with
measurements of the [OIII] $\lambda\lambda$4959,5007 lines,
 translates into an upper limit of 12,700K on the temperature in the
O$^{++}$ zone (assuming $N_e$~=~100~cm$^{-3}$),  and a lower limit on
the oxygen abundance of log(O/H)~$\gtrsim~-$4.07 (both derived using
the FIVEL program (de Robertis, Dufour \& Hunt 1987) as implemented in
IRAF).   As will be shown below, this limit is consistent with the
abundance derived via the semi-empirical method.

\subsection{Semi-Empirical Methods}

Fortunately, in the absence of a reliable [OIII] $\lambda$4363
detection, there exist alternative methods for deriving nebular
abundances which rely on observations of the bright lines alone (eg.
Pagel et al 1979, Skillman 1989, McGaugh 1991, Thurston et al 1996).
Empirical methods to derive the oxygen abundance exploit the
inter-relationship between O/H, T$_e$ and the intensities  of the
strong lines, [OII] $\lambda$3727 and [OIII] $\lambda\lambda$4959,5007,
via the parameter
$$\mathrm R_{23}=\frac{\left[OII\right] \lambda 3727+\left[OIII\right] \lambda\lambda 4959,5007}{H\beta}.$$
As O/H decreases, the cooling efficiency of the nebular gas drops
because there are fewer metal ions, and as a result T$_e$ increases.
This leads to a substantial brightening in the 4959{\AA} and 5007{\AA}
lines, and hence an increase in R$_{23}$.  On the other hand, as O/H
increases, cooling becomes more efficient leading to a decrease in
T$_e$.   Most of the cooling then occurs  through the fine-structure IR
lines at 52{\micron} and 88\micron, leading to a decrease in the strength
of the optical [OIII] lines and therefore in R$_{23}$.  These
variations in [OIII] line strength can clearly be seen in the
representative spectra shown in Figure \ref{refspect}, which are
stacked in order of decreasing metallicity. This simple relationship
between R$_{23}$ and O/H  gets complicated  by the fact that at very
low O/H ($\sim$~30\% solar\footnote{We adopt log(O/H)$_{\sun}=-$3.07
and log(N/O)$_{\sun}=-$0.88 from Anders \& Grevesse (1989).}), the
sheer lack of oxygen causes the  bright lines (and hence R$_{23}$)  to
decrease as O/H decreases, due to the growing importance of Ly$\alpha$
cooling (Edmunds \& Pagel 1984).  As a result, while a single value of
R$_{23}$ can uniquely specify O/H over most of the range in
metallicity, there is a  turnover region (20--50\% solar) where the
relationship becomes double valued.

There have been several calibrations of the O/H--R$_{23}$ relationship
over the past years, based on both observations of HII regions with
known abundances and the results of photoionization models (eg. Edmunds
\& Pagel 1984, MRS, Dopita \& Evans 1986, Skillman 1989, McGaugh 1991,
ZKH).  Of particular importance for the present study is the
calibration at the low abundance end (log(O/H)$<-$3.8), and in the
turnover region ($-$3.4~$\gtrsim$~log(O/H)~$\gtrsim-$3.8) where a
single value of R$_{23}$ corresponds to two values of O/H.  Skillman
(1989) and McGaugh (1991) have illustrated the importance of accounting
for the ionization  state of the nebula in deriving an abundance
estimate at low metallicities, however only the McGaugh calibration
takes explicit account of this  (see McGaugh 1994).  McGaugh's
calibration has the further advantages that with it one can   predict
oxygen abundances on both upper and lower branches of the O/H--R$_{23}$
relation, and also the volume averaged ionization parameter U, defined
as $$\mathrm U=\frac{Q}{4\pi{R_{s}^{2}}Nc}$$  where Q  is the ionizing
photon luminosity, R$_{s}$ is the radius of the Str\"{o}mgren sphere, N
is the  number  density of the gas and c the speed of light.  For these
reasons,  we have adopted this calibration here.

Several methods have been proposed by which to distinguish between
upper and lower branches for objects with values of R$_{23}$ which
place them in the double-valued region.  McGaugh (1994) advocates the
use of the [NII] $\lambda$6584/[OII] $\lambda$3727 ratio, noting that
it varies monotonically with O/H and that it is not very sensitive to
the ionization parameter U since the two ions have similar ionization
potentials.  The division between upper and lower branches of R$_{23}$
is fairly well-defined, with log([NII]/[OII])$>-$1
(reddening--corrected) indicating the upper branch and
log([NII]/[OII])$<-$1 indicating the lower branch.  Another diagnostic
which has been used in the literature is  the value of the line ratio
[OIII]/[NII] (eg. Skillman 1989), with the transition between upper and
lower branches occuring at log([OIII]/[NII])~$\sim$~2.  While this
parameter also varies monotonically with abundance, it is sensitive to
the ionization parameter and is thus of limited use in the low
abundance regime where such effects are important.

\begin{figure}
\plotone{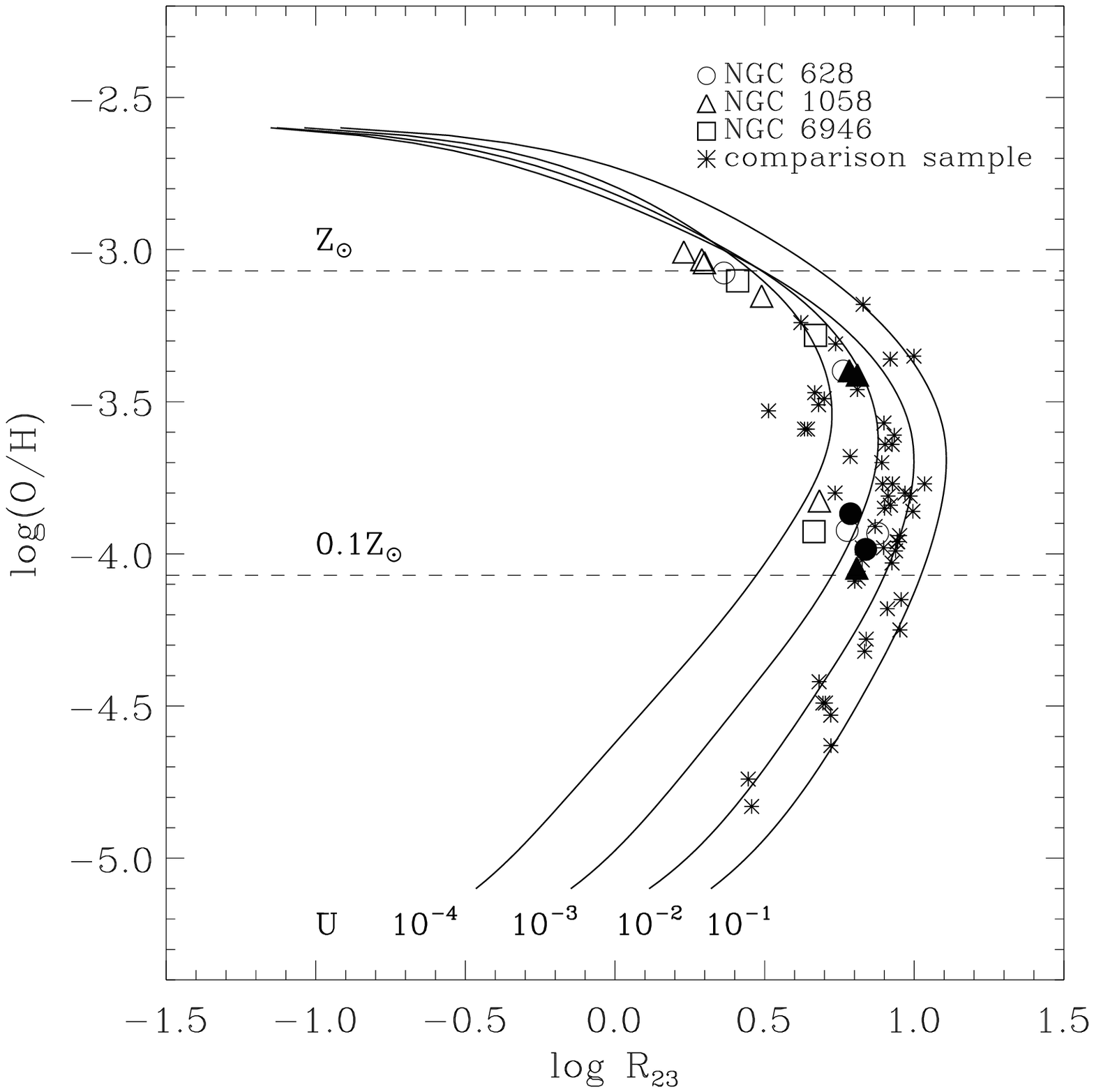}
\caption{Model grid in the log(O/H)--log(R$_{23}$) plane from
McGaugh (1991).  These models have been calculated for an upper mass
cutoff of 60~M$_{\sun}$.  One can see the important effect of the
ionizaton parameter, U, at low abundances.  Overplotted on the grid are
our sample of HII regions, as well as a heterogenous sample of objects
amassed from the literature (see text).  Our HII regions occupy the
same region of the diagram as this comparison sample, for which the
McGaugh calibration is shown to reliably predict abundances (see
Section 5.2).  \label{mg_grid}}
\end{figure}

In Figure \ref{mg_grid}, we plot our HII regions on the model grid of
log(O/H)-- logR$_{23}$ from McGaugh (1991).  One can clearly see the
effect of the ionization parameter, U, in the turnover region and on
the lower branch.  All models converge towards a single upper branch,
reflecting the fact that R$_{23}$ is insensitive to U in this regime.
As can be seen, our outer disk HII regions cluster around the $`$knee'
of the calibration.  We have also overplotted a heterogeneous sample of
HII regions which have oxygen abundances determined in the literature
via the $`$direct' method (see Section 5.2 for a discussion of this
sample).  The objects in this sample populate the same general region
of the diagram as the HII regions in our sample.  As will be discussed
in detail below, comparison of the abundances determined via the
$`$direct' method and via the model calibration for this sample show
very good agreement.  Thus, the McGaugh calibration generally provides
reliable abundance estimates for objects which lie in this region of
the diagram.

Nitrogen to oxygen abundances (N/O)  may be determined in the absence
of  a measurement of the strength of [OIII] $\lambda$4363 by the
algorithm recently proposed  by Thurston et al (1996), again requiring
only observations of the  bright lines [NII],[OII] and [OIII].  The
relationship, which is based on the same premise as that of the
relationship between oxygen abundance and R$_{23}$, is calibrated by
the results of photoionization modelling.  This calibration does not
take account of the fact that, for oxygen abundances lower than
$\sim$~25\% solar, a non-unique relation exists between O/H and
R$_{23}$.

In Table 6, we list the derived oxygen and nitrogen-to-oxygen
abundances for our sample, as well as the mean volume averaged
ionization parameter, derived by our adopted techniques.

\section{Uncertainties}

The uncertainties in the derived metallicities are a combination of
both measurement errors and the intrinsic uncertainties in the model
calibrations.   As we will show below, various tests reveal that the
formal errors on the line strengths appear to  underestimate
considerably the actual uncertainties in  the line ratios, and in the
final derived chemical abundances, and thus are of limited use for
understanding the intrinsic uncertainty in our abundance estimates.

\subsection{Measurement Errors}

Measurement errors can arise from a variety of causes, such as
extraction, sky subtraction,  varying sky conditions and
profile--fitting.  We investigated the magnitude of the measurement
errors in our data by conducting a series of experiments.   First of
all, we compared the line intensity measurements as derived from two
different techniques --  gaussian fitting and direct integration under
the line profile. Defining the $`$mean fractional difference' to be the
result from either method minus the mean, divided by the mean,  we find
values of 1\%~$\pm$~3\%.  While the mean fractional difference was
greatest at the smallest fluxes, it is still less than 10\%.  Next,  we
compared line strength measurements  for those HII regions for which we
had obtained multiple spectra (that were subsequently combined).  Since
multiple observations of a given HII region were always made
consecutively, this comparison should  not be influenced to a great
extent by slit-positioning.
Considering only those lines which are well--detected, we calculate the
mean fractional difference from multiple line measurements to be
8\%~$\pm$~10\%.  Fainter lines tend to show more scatter than strong
lines, for example [OIII] $\lambda$4959 and [NII] $\lambda$6548 have
mean fractional differences of $\sim$~15\%.  A conservative estimate of
the uncertainties in the individual line strengths inferred from this
comparison is $\pm$~10\%, from which we expect uncertainties of
$\sim$~15\% in the line ratios.

We also calculated the mean values of the ratios [OIII]
$\lambda$5007/[OIII] $\lambda$4959 and [NII] $\lambda$6854/[NII]
$\lambda$6548 for the HII regions in our sample.  Since these ratios
have a fixed theoretical value, comparison of our sample average with
the expected values provides an additional gauge of our measurement
errors.  Considering again only those lines which are well--detected,
we find $<$[OIII] $\lambda$5007/[OIII] $\lambda$4959$>=$ 2.88 with
standard deviation 0.31 and $<$[NII] $\lambda$6854/[NII]
$\lambda$6548$>=$ 2.96~$\pm$0.39, which compare favourably to the
theoretical values of 2.88 (Nussbaumer \& Storey 1981) and 2.95
(Mendoza \& Zeippen 1982) respectively.

As a final  check on measurement errors, we included in our sample two
HII regions which were observed by MRS.  Despite the likely differences
in pointings and aperture sizes employed, we find an excellent
agreement in the values of R$_{23}$; $\Delta$~log(R$_{23}$)=0.144  for
FGW 628--A (MRS identification N0628(-074,-022)) and $-$0.068 for FGW 6946--A
(MRS identification N6946(+182,+103)).  These differences in R$_{23}$
lead to differences of only $-$0.08 and 0.07 dex in log(O/H), and
$-$0.16 and 0.08 in log(N/O).

In summary, these various tests reveal that our measurement errors are
small, and we conservatively estimate line ratios to be accurate to
better than 15\%, or $\sim$0.1 dex.

\subsection{Calibration Errors}

The dominant source of uncertainty in our results is without a doubt
that  due to the  model calibrations of the semi-empirical
relationships between line strength and elemental abundance.
Propagating the formal errors on the line strengths through the
equations used to derive the abundances produces formal errors on the
metallicities of only $\sim$~0.05 dex. As we will discuss below, the
uncertainty in the absolute value of the calibration is likely to be
much  larger than this.

Uncertainties in the calibration arise from the limitations in the
inputs to the models used to construct the calibration, and from the
ability of the calibration to reproduce model input data, and indeed to
reproduce the values of chemical  abundances which have been determined
via a measurement of the electron temperature.   As discussed by
McGaugh (1994), while these uncertainties can have an important effect
on the  absolute metallicities derived, they have a much smaller impact
on the  relative values.

One of the most important input parameters is the  shape of the
ionizing spectrum, which depends on both the mass and the metallicity
of the ionizing stars (McGaugh 1991).  McGaugh (1991) treats this
problem by assuming a cluster containing several tens of OB stars is
responsible for the ionization, which when averaged over, produces a
constant ionizing spectrum, relatively insensitive to both the
metallicities and individual effective temperatures of the enclosed
stars, and hence the IMF.   We note however that many of our outer disk
HII regions, if ionization-bounded, are consistent with ionization by
only a few massive stars (see Table 2) and hence the assumptions which
have gone into  McGaugh's calibration may not be entirely appropriate
for the objects under study here.

Concerning the ability of the calibrations to reproduce model input
data, both McGaugh(1994) and Thurston et al (1996) report relative
uncertainties  of 0.1--0.2 dex over a wide range in metallicities.  In
the turnover region where R$_{23}$ is double-valued,  McGaugh (1994)
estimates uncertainties of $\gtrsim$ 0.2 dex, although as he points out
they cannot be too much larger than this since strong R$_{23}$
guarantees that an HII region lies in the range
--3.4~$\ge$~log(O/H)~$\ge$--3.8.   Thurston et al (1996) see increasing
deviations from the output, relative to model input, at low
metallicities, but  their conclusion is based on tests with only two
low metallicity models:  at 30\% solar the deviation is $<$ 0.1 dex
whereas it is  $\sim$0.3 dex  for the lowest metallicity model tested,
which has an oxygen abundance of 7\% solar.

Perhaps the most robust measure of the accuracy of the semi-empirical
technique is a direct comparison of predicted abundances with those
measured for low metallicity HII regions which have published abundance
determinations based on a measurement of the electron temperature.   We
have gathered from the literature a sample of low metallicity HII
regions, in a heterogeneous set of parent galaxies, ranging from low
metallicity dwarfs, including blue compact dwarfs (eg. Izotov et al
1994, Skillman \& Kennicutt 1993, Skillman et al 1994, Miller 1994),
irregulars (Webster \& Smith 1983, Pagel et al 1980, Miller 1994) and
spirals (Webster \& Smith 1983, Edmunds \& Pagel 1984, Pagel et al
1979, 1980, Garnett et al 1997a, Vilchez et al 1988).  The sample also
includes two  outlying HII regions in the spirals M81 and M101 (Garnett
\& Shields 1987, Garnett \& Kennicutt 1994); the M81 HII region is
comparable in luminosity to those outer disk HII regions in our sample,
having an H$\alpha$ luminosity of 4~$\times$~10$^{37}$~erg~s$^{-1}$,
while the M101 HII region is significantly more luminous.

\begin{figure}
\plotone{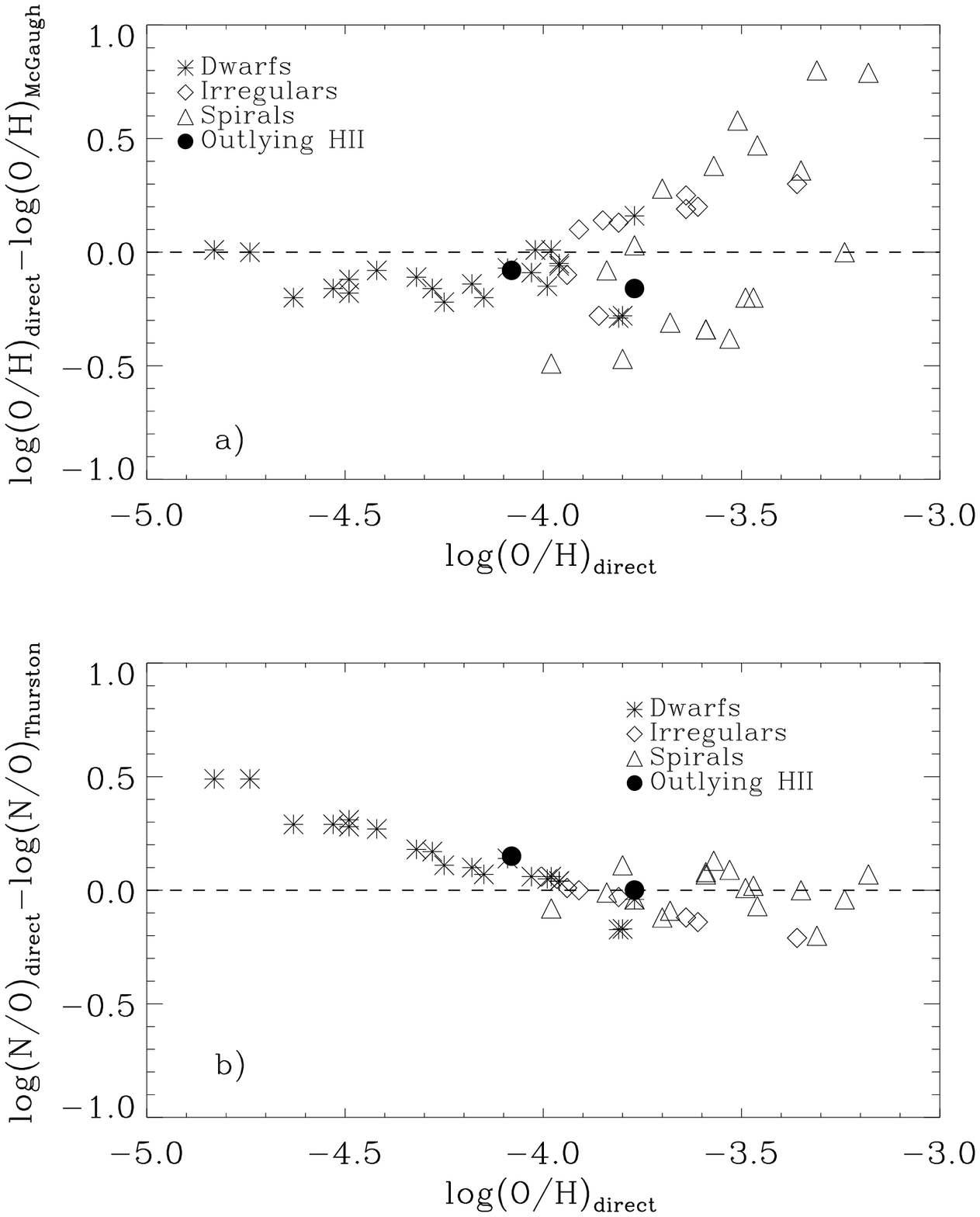}
\caption{Comparison of O/H  (a) and N/O (b) abundances
determined from the $`$direct' and semi-empirical techniques, expressed
as a function of oxygen abundance.  The data points represent a
hetereogenous sample of low metallicity objects as described in the
text.  Over the prime region of interest for outer disk HII regions
(10-30\% solar), we find very good agreement between
empirically-determined abundances and those determined from a measure
of the electron temperature.  \label{comp}}
\end{figure}

The top panel of Figure \ref{comp} shows the difference between the
$`$direct' abundance determination (using the derived value of T$_e$)
and that which is returned by the McGaugh calibration based on only the
bright line strengths.  As can be seen, the agreement between the two
techniques improves significantly with decreasing abundance.  While the
larger discrepancy between the two techniques at high abundances may be
due in part to shortcomings of the McGaugh calibration (for example,
the lack of account for dust and depletion of heavy elements (see
Shields \& Kennicutt 1995)), it also reflects the increasing difficulty
to measure accurate [OIII] $\lambda$4363 strengths, and hence
abundances via the $`$direct' method, at metallicities close to
solar\footnote{For those HII regions which clearly lie on the upper
branch, we also calculated what the abundance would be if we used the
R$_{23}$ calibration of ZKH.  We find an average offset of
0.13~$\pm$~0.06 dex between the different determinations of log(O/H),
which is consistent with the uncertainty produced by either method
alone.}.  We confirm the effect noted by McGaugh (1991) that the model
calibration tends to slightly overpredict O/H at very low abundances.
Still, the agreement between the independent determinations is
generally very good, with an average offset of only --0.02 dex, and a
standard deviation of 0.28 dex across the entire range of abundances
spanned by our comparison sample.  Over the particular region of
interest for outer disk HII regions (10--30\% solar), we find a mean
offset between the different methods of --0.06~$\pm$~0.23 dex.  Hence,
we will adopt 0.2 dex as the typical uncertainty in our derived oxygen
abundances.

The bottom panel of Figure \ref{comp} shows the same comparison for the
nitrogen-to-oxygen abundances determined via the Thurston et al (1996)
calibration.   As can be seen, the Thurston calibration provides N/O
abundances which are in excellent agreement with those measured
directly, deviating significantly only at extremely low metallicities.
Averaging over all metallicities, we find a mean offset of 0.06 dex
with a standard deviation of 0.16 dex. Over our prime region of
interest (10--30\% solar), the mean offset between the $`$direct'
abundance and model predictions is only --0.03 dex and the dispersion
0.08 dex.  We also  confirm their noted trend of the model
under-predicting true abundances at very low O/H; systematic deviations
of $\gtrsim$0.3 dex appear for metallicities less than 5\% solar.
 Thus, the semi-empirical techniques adopted in the present work appear
 to be able to reproduce (surprisingly) well the chemical abundances in
this heterogeneous sample of low metallicity objects.  In the
discussion that follows, we will  thus  adopt a $\pm$~0.2 dex
uncertainty in our oxygen abundances and $\pm$~0.1 dex in
nitrogen-to-oxygen abundances.
 
\section{Results}  

\subsection{Electron Densities}

\begin{figure}
\plotfiddle{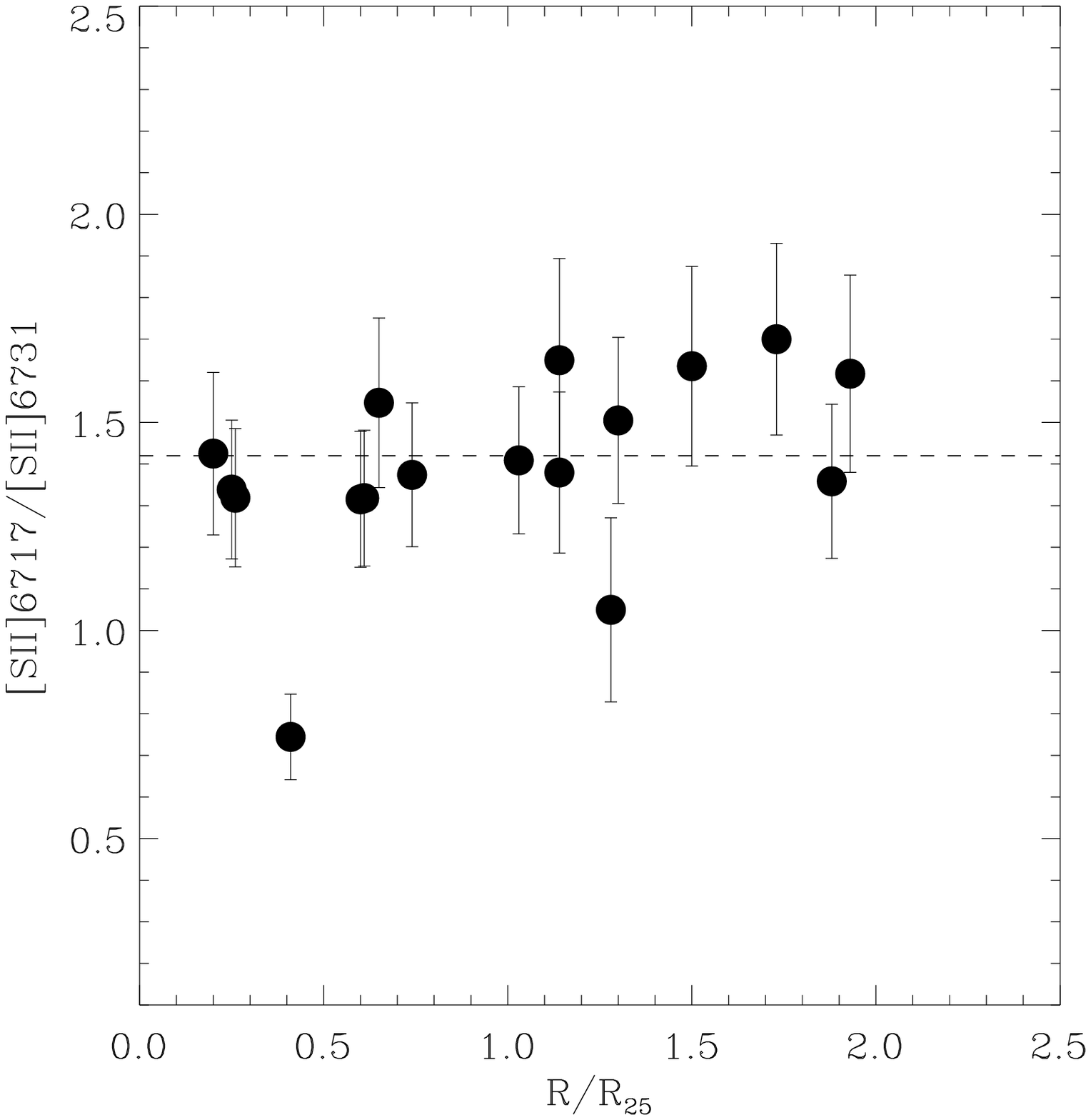}{90truemm}{0}{70}{70}{-220}{0}
\caption{Variation of the density sensitive ratio, [SII]
$\lambda$6717/[SII] $\lambda$6731, for our entire sample of HII regions
plotted as a function of radius, normalised to the size of the optical
disk.  The dashed line indicates the low density limit.   \label{dens}}
\end{figure}

Electron densities can be estimated via the ratio of the [SII]
$\lambda\lambda$6717,6731 lines (eg. Osterbrock 1989).  The derived
ratios for our entire sample are plotted in Figure \ref{dens} as a
function of deprojected radius, normalised to the optical radius.  The
horizontal line indicates the low density limit of 1.42 (Czyak et al
1986); HII regions with ratios comparable to this value are not
affected by collisional de-excitation while those with lower values
are.  As can be seen, most of our sample appears consistent with the
low density limit and hence we infer electron densities
$\lesssim$~100~cm$^{-3}$.   There is one notable exception, FGW 1058--C,
for which we infer a density of $\sim$10$^3$~cm$^{-3}$.  No evidence
exists for trends in electron density with galactocentric radius.

\subsection{Ionization Parameter}

\begin{figure}
\plotfiddle{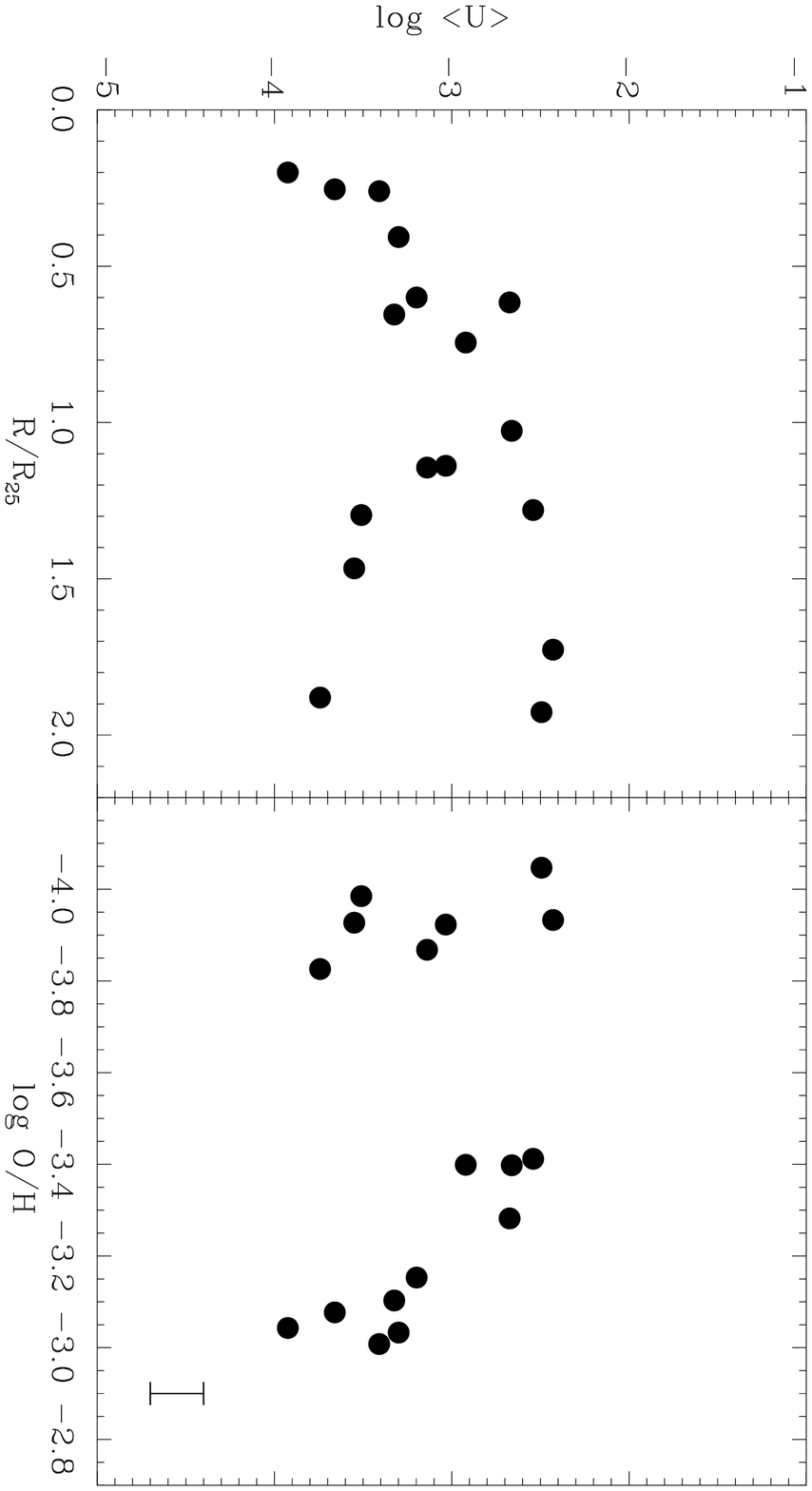}{90truemm}{90}{70}{70}{250}{0}
\caption{Variation of the volume-averaged ionization parameter,
$<$U$>$, for our entire sample of HII regions, plotted as a function of
radius (right), normalised to the size of the optical disk, and as a
function of oxygen abundance (left).  The bar in the lower right corner
indicates the probable error on these determinations.  \label{ion} }
\end{figure}

The ionization parameter of a nebula, previously defined in Section
4.2, is essentially the local ratio of Lyman--continuum photons to gas
density, which determines the degree of ionization at any particular
location within the nebula.   Figure \ref{ion} shows the variation of
the mean volume averaged ionization parameter, $<$U$>$, as derived by
the McGaugh (1994) calibration, as a function of both galactocentric
radius and oxygen abundance.  HII regions at large radii are observed
to exhibit a large range in $<$U$>$, and there is no obvious trend in
ionization parameter with either radius or oxygen abundance.  This
latter finding is in agreement with the results of ZKH and Kennicutt \&
Garnett (1996), but contrary to  Evans \& Dopita (1985) who have
proposed that U is anti-correlated with abundance.
 
\subsection{Reddening and Extinction}

\begin{figure}
\plotone{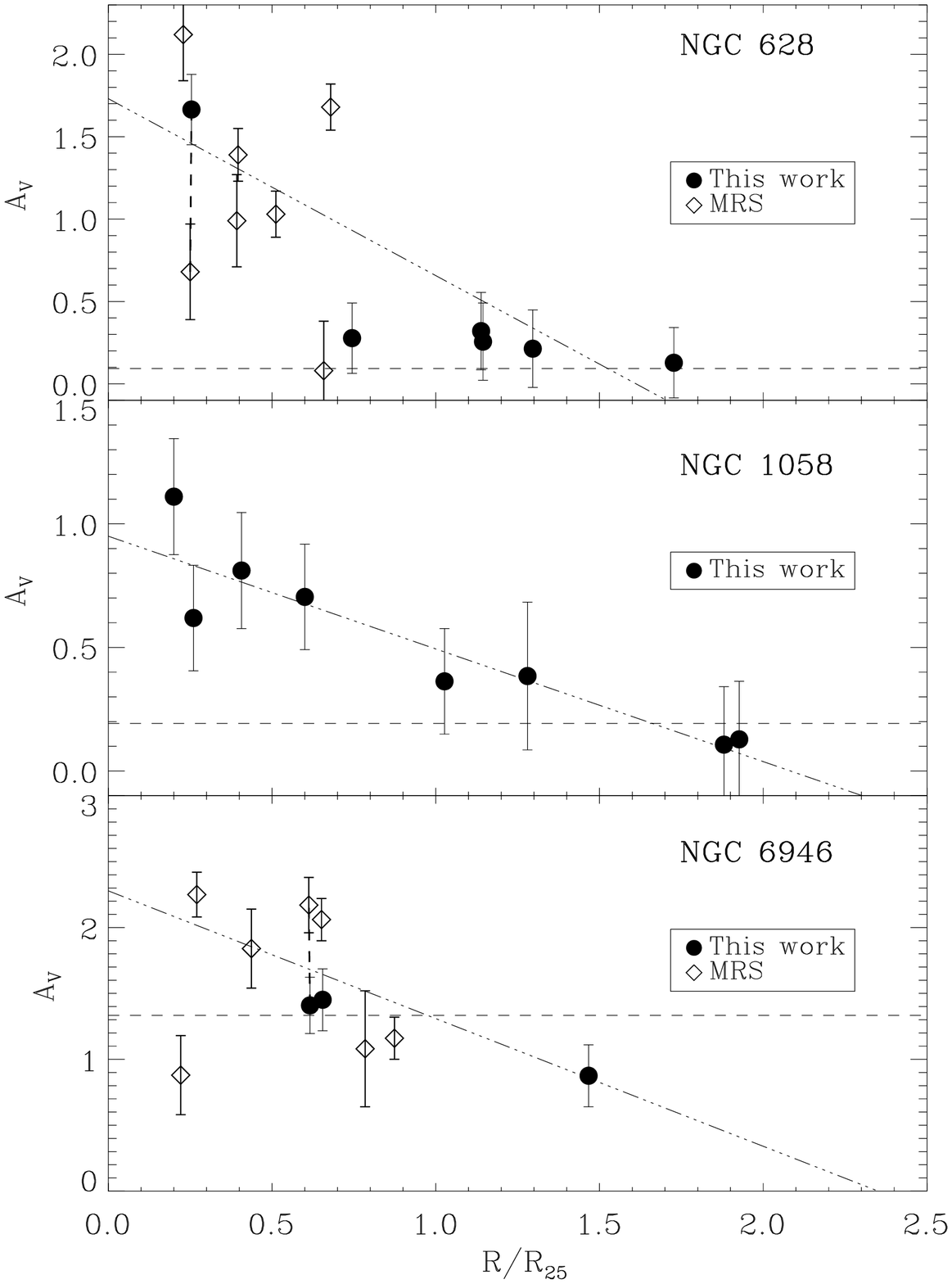}
\caption{Radial variation of extinction for our galaxy sample,
expressed in terms of the optical radius R$_{25}$.  Our data points are
indicated by filled circles, those from MRS are indicated by open
diamonds.  The horizontal dashed line indicates the level of Galactic
extinction towards these galaxies, taken from the RC3.  The
dashed-dotted line indicates a linear least squares fit the points.
Thick dashed lines join independent measurements of the same HII
region.\label{ext}}
\end{figure}

Figure \ref{ext} shows the radial variation of the extinction for each
galaxy, as derived from the reddening of the Balmer decrement.  Also
shown are the extinction measures from MRS for their inner disk HII
regions.  A large dispersion in extinction can be seen over the face of
these galaxies, clearly indicating the dominant effect of local
variations over radial variations within the optical disk.  Our derived
extinctions for the two HII regions in common with MRS show deviations
of $\sim$ 1 magnitude, which may  partly reflect variations in
extinction over very small scales, accountable for by differences in
pointing alone. Despite the large scatter which typifies the inner
disk,  clear evidence exists for diminished extinction at large
radii.    When account is made for the Galactic extinction towards
these galaxies (indicated by the horizontal dashed line in Figure
\ref{ext}), the outer HII regions are consistent with internal
extinctions of only A$_V$~$\sim$~0--0.2 magnitudes.

We have used a weighted linear least-squares algorithm to fit the
radial behaviour of the extinction in each galaxy and find
 A$_V=$1.73$-$1.07R/\ropt(NGC~628), 0.95$-$0.46R/\ropt (NGC~1058) and
2.28$-$0.97R/\ropt (NGC~6946).  Although clearly not an appropriate way
to characterize the actual extinction in galactic disks, these
parametrizations serve as a means to compare the global extinction
properties of different galaxies.  As can be seen, both the central
value  and the amplitudes of the gradients vary considerably from
galaxy to galaxy. A radial gradient in extinction has also recently
been detected  in M101 by Kennicutt \& Garnett (1996), but most
previous studies have found evidence for only a very weak radial
dependence.  Our results suggest that this may be due in part to the
limited radial coverage of such studies, which typically have sampled
only the inner parts  of galaxies where the amount of scatter dominates
over any existing radial trend (eg. ZKH; MRS;  Belley \& Roy 1992;
Scowen et al 1992).  Interestingly, the radial gradients in A$_V$ found
here are comparable in magnitude to those found for the radial
variation of the oxygen abundance (see below), lending support for the
idea that the changes in A$_V$ are driven largely by a decrease in
available metals (see Table 7).

\subsection{Oxygen}

\begin{figure}
\plotone{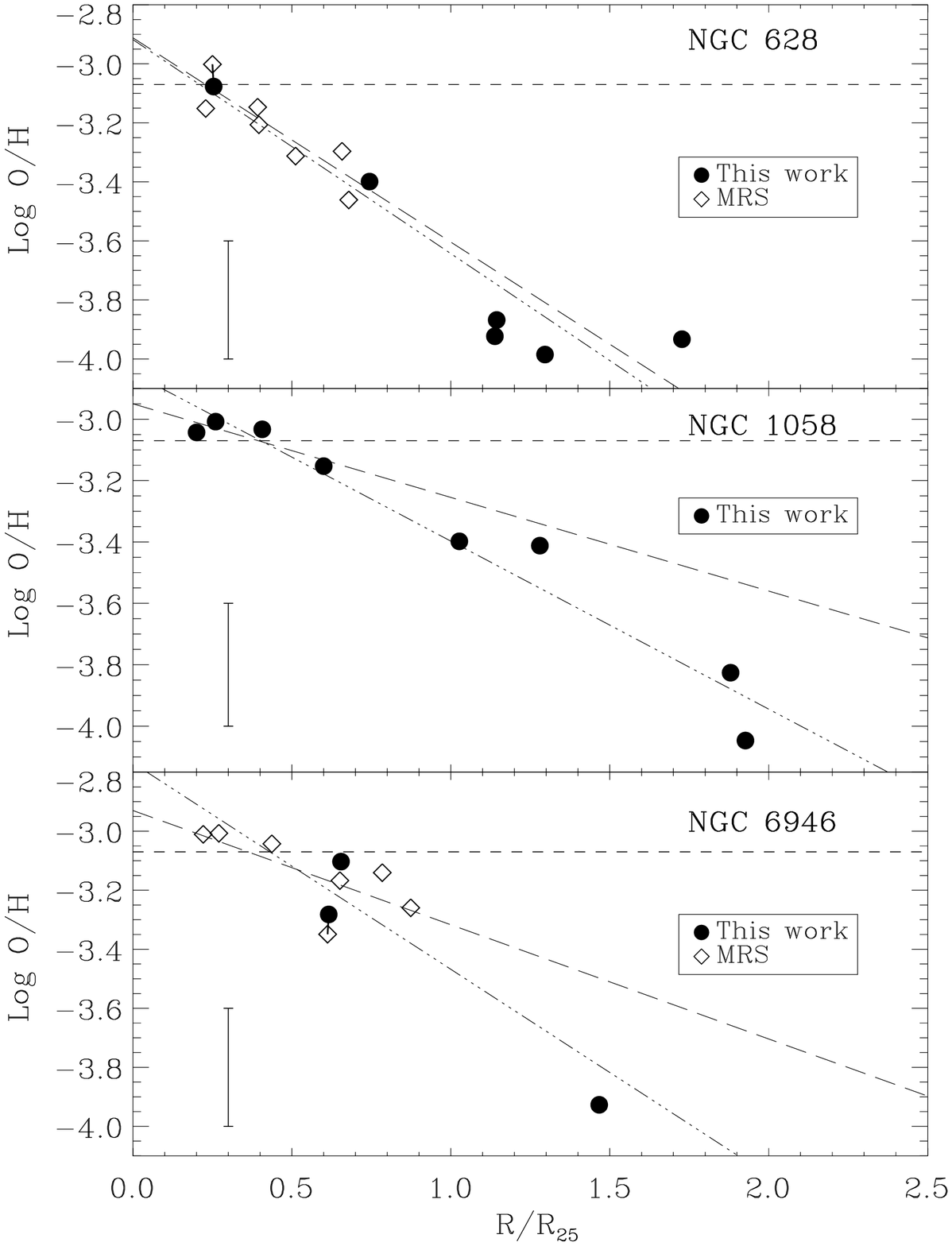}
\caption{Radial variation of the oxygen abundance for our galaxy
sample, expressed in terms of the optical radius R$_{25}$. Our
abundance determinations are indicated by filled circles, and those
taken from MRS are indicated by open diamonds.  The horizontal dashed
line indicates the solar abundance (Anders \& Grevesse 1989).  The
dashed-dotted line indicates a linear least squares fit to all points,
whereas the long dashed line indicates the fit to only those points
lying within the optical radius. Solid lines join repeat measurements
of the same HII region.  The error bar in lower left indicates probable
error due to calibration uncertainties.\label{oxy}}
\end{figure}

In Figure \ref{oxy}, we present the derived O/H abundances for the
three galaxies in our sample as a function of the deprojected
galactocentric radius, normalised to the optical radius.  Since we have
demonstrated that the dominant errors in the abundance determinations
arise from the use of model calibrations and not from the measurements
themselves, we omit placing formal error bars on the data points and
instead place a representative  error bar in each corner of the plot to
indicate the estimated calibration uncertainty ($\pm$~0.2~dex) as
derived above.  Also shown are the abundances we derived using our
present technique for the inner disk HII regions observed in NGC~628
and NGC~6946 by MRS (calculated from their published
reddening-corrected line strengths).    Our repeat measurements of the
two MRS inner disk HII regions are indicated by a solid line which
joins the independent measurements.   As can be seen, there is an
excellent agreement between these determinations.

Inner disk abundances typically range from 20\% above solar to 40\%
below solar while beyond the edge of the optical disk, the measured
oxygen abundances are $\sim$~10--50\% solar.
 The outermost abundances in all three galaxies are $\sim$~10-15\%
solar, measured at radii in the range 1.5--2~R$_{25}$.   We have used a
linear (uniformly-weighted) least squares routine to fit the variation
of O/H as a function of deprojected galactocentric radius, expressed in
terms of both the optical radius and in terms of kiloparsecs (see Table
7).  We have also fit the gradients using only those points lying at or
within the optical radius, in order to gauge the importance of the
outer disk measurements in defining the abundance gradient.  In this
case, we find log(O/H)=$-$2.91$-$0.69($\pm$0.12)R/\ropt (NGC~628),
$-$2.95$-$0.30($\pm$0.14)R/\ropt (NGC~1058) and
$-$2.93$-$0.39($\pm$0.16)R/\ropt (NGC~6946).  These gradients are
consistent with those derived for NGC~628 and NGC~6946 by ZKH, which
were determined using weighted linear least-squres fits to only the
inner disk samples of MRS and employing a different calibration of the
O/H--R$_{23}$ relationship than that which is adopted here.  For
comparison, they find gradients in units of the optical radius of
$-$0.96~($\pm$~0.32) for NGC~628 and $-$0.55~($\pm$~0.26) for
NGC~6946.  As is evident, the values of the outer disk abundances play
a crucial role in defining the abundance gradient across the disk.
Consideration of only inner disk abundances results in significantly
flatter gradients  for NGC~1058 and NGC~6946.

\begin{figure}
\plotone{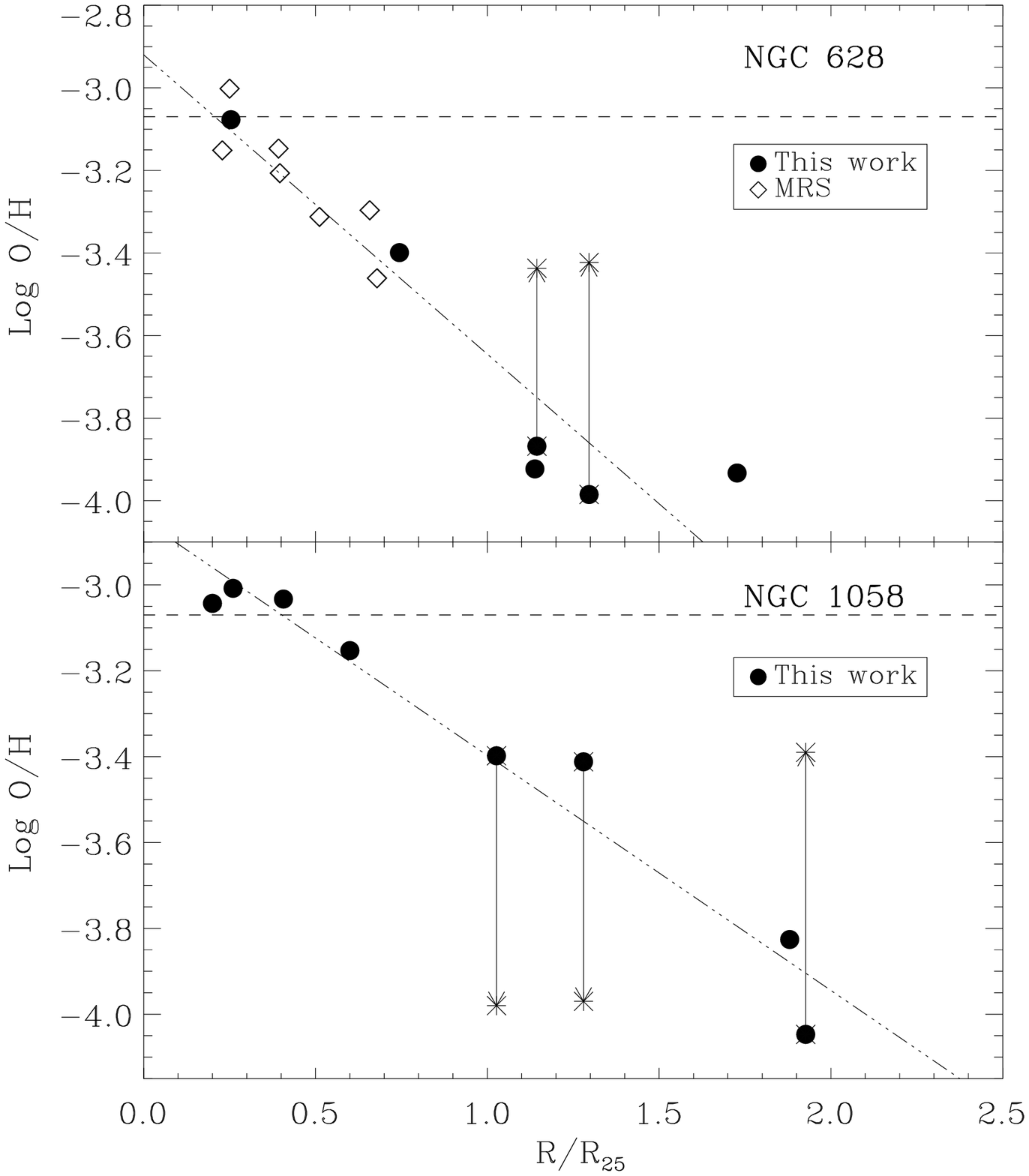}
\caption{As in  Figure \ref{oxy}, except we have explicitly 
indicated the change to the observed abundance gradients if
the opposing branch is chosen for those HII regions where the
placement is uncertain.   The asterixes mark the new abundances
for the HII regions in question.   \label{oxy2}}
\end{figure}

As we have discussed, a crucial aspect of the abundance determination
via the semi-empirical method is determining whether a given HII region
lies on the upper or lower branch of the calibration.  Several of the
extreme outer disk HII regions in our sample have values of
log([NII]/[OII]) close to unity, which makes the branch placement
uncertain. In Figure \ref{oxy2}, we show the effect of choosing the
other branch in those cases in which the branch placement is somewhat
ambiguous (defined here to be those cases where log([NII]/[OII]) lies
within $\pm$~0.1 dex of -1).  Only NGC~628 and NGC~1058 have HII
regions which fall in this category.   As can be seen, choice of the
opposing branch for these HII regions can have a profound effect on the
nature of the gradient, however it leads to rather strange (and
unlikely) behaviours at large radii.  We intend  to obtain measures of
[OIII] $\lambda$4363 line, and thus the oxygen abundance via the
$`$direct' method, for those HII regions in which the abundance
determination from the semi-empirical method is uncertain.

\subsection{Nitrogen-to-Oxygen}

\begin{figure}
\plotone{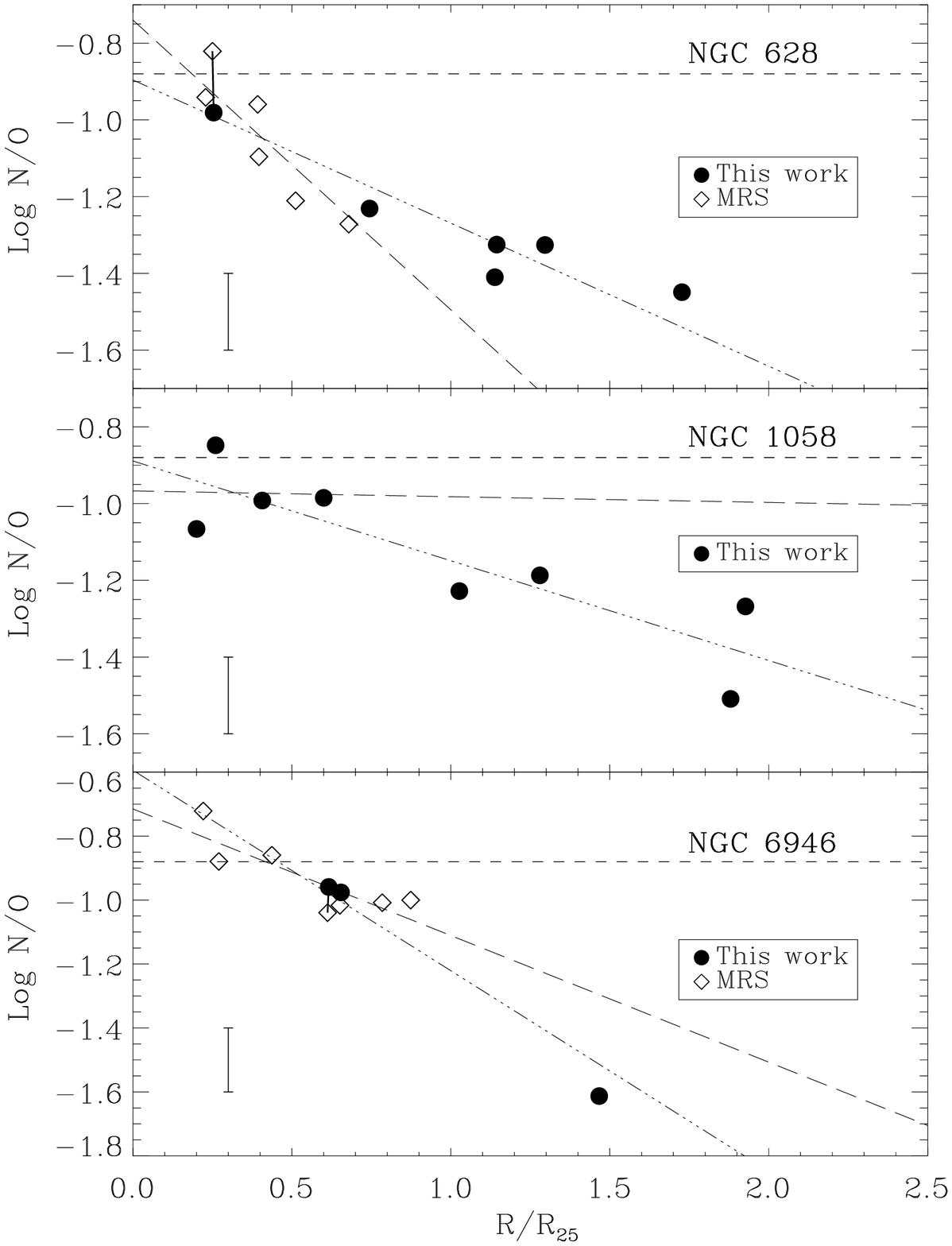}
\caption{Radial variation of the nitrogen-to-oxygen abundance for
our galaxy sample, expressed in terms of the optical radius R$_{25}$.
Our abundances  are indicated by filled circles, whereas those derived
from MRS are indicated by open diamonds.  The horizontal dashed line
indicates the solar abundance (Anders \& Grevesse 1989).  The
dashed-dotted line indicates a linear least squares fit to all points,
whereas the long dashed line indicates the fit to only those points
lying within the optical radius. Solid lines join repeat measurements
of the same HII region.  The error bar in the lower  left indicates the
probable error due to calibration uncertainties.\label{nitro}}
\end{figure}

Nitrogen-to-oxygen abundance ratios range from 50\% above solar to
40\%  below solar across the inner disks, and from 20-50\% solar beyond
the optical radius (see Figure \ref{nitro}).  The outermost regions
have typical abundances of 20--25\% solar.  Linear least-squares
(uniformly-weighted) fits have been carried out to characterize the
radial behaviour (see Table 7).  Fitting only those points lying at or
within \ropt produces gradients of
log(N/O)=$-$0.74$-$0.76($\pm$0.13)R/\ropt (NGC~628),
$-$0.97$-$0.02($\pm$0.36)R/\ropt (NGC~1058) and
$-$0.72$-$0.40($\pm$0.09)R/\ropt (NGC~6946).   Once again, it can be
seen that the outermost abundances play a key role in defining the
abundance gradient.  In particular, one might deduce no gradient in N/O
if presented with only the inner disk measurements, whereas the outer
disk measurements clearly show a decline in N/O at large radii.   

\section{Discussion}

\subsection{Constraining Outer Galactic Abundance Gradients}

The results presented here constitute the largest set of extreme outer
disk abundances ever measured, and probe the chemical abundance content
of these optically faint, previously unexplored regions of present day
galactic disks.   We have found that the outermost HII regions studied
have O/H abundances in the range  of 10--15\% solar and  N/O abundances
in the range of 20--25\% solar.  Such low abundances have rarely
been measured before in spiral disks (eg. Garnett \& Kennicutt
1994).

We have shown that the outermost abundances play an important role in
defining abundance gradients across the disks, and indeed often change
the  nature of the gradient.  For example, in two of the galaxies
studied here, the oxygen gradients are observed to steepen considerably
when account is taken of the outermost HII regions.  Within the limits
of the current dataset, the radial abundance gradients are consistent
with single log-linear relationships, although hints of interesting
behaviour can been seen at large radii in two of the galaxies.  There
could be a flattening of the oxygen abundance beyond the edge of the
optical disk in NGC~628 (although this result largely hinges on the
metallicity of the outermost HII region), as well as a  steepening of
the outer gradient in NGC~1058.  Unfortunately, given the relatively
large errors in our abundance determinations, as well as the relatively
small size of the current sample, it is not yet possible to assess the
significance of these features.   Similarly, it is not yet possible to
constrain the amount of intrinsic scatter in the abundances as a
function of galactocentric radius, which is of interest for
understanding the efficiency and timescales for elemental mixing  at large
radii.  While the two outermost HII regions in NGC~1058 show
differences of $\gtrsim$ 0.2 dex in both O/H and N/O despite lying at
very similar radii, a knowledge of the electron temperatures of these
regions will be required before we can distinguish between real scatter
and uncertainties inherent in the model calibrations.

Chemical evolution models predict a variety of different behaviours for
galactic abundance gradients (eg. Prantzos \& Aubert 1995, Molla et al
1996, Wyse \& Silk 1989, Clarke 1989).   Of these models, some predict
simple exponential declines while others produce steepenings or
flattenings in the outer disk.    Our current dataset is not yet
sufficient to discriminate between various chemical evolution models,
but  we are continuing to obtain more and better data on these and
other galaxies, spanning a range of physical environments and Hubble
types.

\subsection{Comparison to Other Galaxies}

It is of great interest to compare our measurements of outer disk
abundances with the few published measurements in the literature.  High
quality measurements exist for the Galaxy (Fich \& Silkey 1991; Vilchez
\& Esteban 1996;  Rudolph et al 1997; Afflerbach et al 1997), M81
(Garnett \& Shields 1987) and M101 (Garnett \& Kennicutt 1994), but
these measurements do not extend to the extreme galactocentric radii
that we have studied here.

In case of Galaxy, outer disk HII regions have been  studied via
optical and far-IR techniques out to 1.3$R_{edge}$, where  R$_{edge}$,
the edge of the optical stellar disk, is taken to be 14~kpc (Ruphy et
al 1997)\footnote{It remains unclear as to the relationship between the
optical edge, as defined by star counts, and the 25th B-magnitude
isophote.}.  Both Fich \& Silkey (1991) and Vilchez \& Esteban (1996)
found evidence for flat gradients in the nitrogen abundance beyond the
solar circle; additionally, Vilchez \& Esteban (1996) found evidence
for only a mild outer gradient in oxygen abundance, with outer values
of $\sim$~20\% solar, and outer disk values of N/O consistent with
those measured in the solar neighbourhood.  More recently, Rudolph et
al (1997) have re-examined outer abundance gradients using FIR lines
for a sample of 5 HII regions.  When combined with the results of other
studies, these authors do not see compelling evidence for a flattening
of  the outer abundance gradients, and are able to fit the available
data  with single log-linear relationships.  They caution, however,
that their present sample is not yet sufficient to rule out the
possible existence of a flattening  of the abundance gradient in the
outer Galaxy.   They also find the variation of N/O throughout the
Galactic disk to be consistent with a step function, with a mean
log(N/O) of $-$0.50~$\pm$~0.02 for R$<$6.2 kpc and, $-$0.83~$\pm$~0.04
for R$>$6.2 kpc.  We note that HII regions have been discovered at much
larger Galactocentric radii, out to $\sim$ 28 kpc or 2R$_{edge}$ (de
Geus et al 1993) however no abundance determinations have been made for
these as of yet.

In M81, the outermost HII region (lying at $\sim$ 1.3~R$_{25}$) has an
O/H abundance of roughly 20\% solar, consistent with an extrapolation
of the inner gradient, whereas the N/O abundance is close to solar and
consistent with there being no gradient across the disk (Garnett \&
Shields 1987). On the other hand,  the outermost HII region studied in
M101 ($\sim$ 1.1~R$_{25}$) has an O/H abundance of only 10\% solar and
an N/O abundance of 25\% solar; both N/O and O/H are observed to
decrease more or less smoothly with increasing galactocentric radius
across the disk (Garnett \& Kennicutt 1994).  Curiously, these
abundances are as low as those observed in the extremeties of the
galaxies studied here, despite that fact that we have probed the gas at
significantly larger distances beyond the optical disk.

We find that the galaxies studied in the present work are more akin to
M101 than the Galaxy or M81 since the outermost abundances of both O/H
and N/O  are observed to decrease more or less smoothly across their
disks.   The striking difference between the level of N/O enrichment
seen in the outer disks of galaxies studied here (and M101)  and in the
outer disks of the Galaxy and M81 is particularly puzzling and clearly
warrants further study.

\subsection{Implications for Understanding the Evolution of Galactic Disks}

Our outer disk measurements are largely consistent with those observed
in  other low gas surface density objects, such as gas-rich dwarf
irregulars (eg. Garnett 1990; Skillman et al 1989; Skillman et al 1997;
Thuan et al 1995; Miller \& Hodge 1996)\footnote{Note however that the
measured outer disk metallicities are still considerably in excess of
the most metal poor, gas-rich objects known locally, eg.
IZw18, UGC4483 and SBS 0335-052. These objects  have O/H abundances of
only 2--3\% solar (eg. Skillman \& Kennicutt 1993, Skillman et al 1994,
Izotov et al 1997), even although there is evidence for moderately long
periods ($\gtrsim$~10~Myr) of star formation in at least one case
(eg.  Garnett et al 1997b).}  and some low surface brightness galaxies
(McGaugh 1994).  It has often been suggested that outer galactic disks
are $`$built' up through the accretion of gas, either smoothly (eg.
Gunn \& Gott 1972; Larson 1976) or as gas-rich low mass companions (eg.
White \& Rees 1987; Kauffman et al 1993; Kamphuis 1993; Zaritsky
1995).  There would appear to be no obvious argument against this
hypothesis on the basis of chemical abundance content alone.

On the other hand, one might consider the scenario in which outer
galactic disks evolve in relative isolation, with little  inflow or
outflow.  The simple $`$closed box' model of Schmidt (1963) can be used
to predict the mean metallicity expected for such regions, under the
assumption of instantaneous recycling (appropriate for oxygen).   The
closed box model can be represented by a simple relation (Searle \&
Sargent (1972)) $$\mathrm Z=-p~ln~\mu $$ where  p is the yield of the
element in question and $\mu$ is the gas fraction, defined as baryonic
mass in gas to the total baryonic mass (stars + gas).  Our deep B-band
surface photometry yields B-band surface brightnesses at two optical
radii of $\sim$~0.1~L$_{\sun}$~pc$^{-2}$ in NGC~628 and NGC~6946 and
$\sim$~1.2~L$_{\sun}$~pc$^{-2}$ in NGC~1058 (Ferguson et al 1998b).
Under the assumption of M/L~$\sim$~2, this leads to surface mass
densities of 0.2 and 2.4~M$_{\sun}$~pc$^{-2}$.  The B-band M/L ratio is
sensitive to the star formation history of the outer disk, but is
unlikely to be too much larger than the adopted value, which is found
for the solar neighbourhood (Kuijken \& Gilmore 1989).  Inspection of
the HI maps of these galaxies reveals typical HI surface densities of
$\sim$~1~M$_{\sun}$~pc$^{-2}$ (NGC~628) to
$\sim$~3~M$_{\sun}$~pc$^{-2}$ (NGC~1058, NGC~6946) in the extreme outer
disks (van der Kruit \& Shostak 1984, Shostak \& van der Kruit 1984,
Kamphuis 1993).  Correcting for helium (a factor of 1.3), we derive gas
fractions of $\sim$~0.6 in NGC~1058 and $\sim$~0.90 in NGC~628,
NGC~6946.  Adopting a yield of 0.5~Z$_{\sun}$, consistent with the
observed mean metallicity of stars in the solar neighbourhood (Wyse \&
Gilmore 1995), the simple model then predicts  oxygen abundances of 6\%
solar for NGC~628, 25\% solar for NGC~1058 and 2\% solar for NGC~6946.
These values can be compared to the observed metallicities of 10--15\%
solar.  Despite the many uncertainties involved, both NGC~628 and
NGC~1058 have metallicities which lie within a factor of two of the
closed--box model predictions, thus suggesting that the role of gas
flows in the evolution of their outer disks is similar to that for the
solar neighbourhood.  The outer disk of NGC~6946, on the other hand,
does not appear to fit this picture. In the future, we plan to investigate the
predictions of the simple model in more detail, including study of
the radial variation in the derived effective yield.

\subsection{Comparison to High Redshift Damped Lyman-$\alpha$ Systems}

Much recent attention has been focused on uncovering the nature of the
damped Ly$\alpha$ systems (DLAs)  which cause absorption in the spectra
of high redshift quasars.  While there remains much debate on this
topic, follow-up imaging and spectroscopy of these systems lends
support for the idea that at least some of them are young disk galaxies
(eg. Wolfe 1988;  Briggs et al 1989;  Djorgovski et al 1996).  The
distribution of impact parameters for these systems is not
well-defined; however, in the particular case of Djorgovski et al
(1996), an impact parameter of 18~kpc is inferred which clearly places
the line-of-sight in the outer parts of what appears to be a large disk
galaxy at z$=$3.15.   How do the chemical abundances measured in the
DLAs compare to those measured in the extended parts of local
disk galaxies?  There have been many recent efforts to measure the
gas-phase chemical abundances in DLA systems (eg. Pettini et al 1994,
1995, 1997; Wolfe et al 1995; Lu et al 1996, 1998); studies have
generally found them  to exhibit a wide range of low metallicities,
ranging from less than 1/100th of the solar value to 1/10th solar, and
showing a trend of increasing metallicity with decreasing redshift,
albeit with considerable scatter at all redshifts (eg. Lu et al 1996,
Pettini et al 1997).   Pettini et al (1997) calculate a column
density-weighted mean metallicity of 1/13th solar for their sample of
34 DLAs (0.7$<$z$<$3.4), using the Zn lines which should be relatively
unaffected by depletion onto dust.\footnote{Note that the average HI
column for this sample of DLA systems is roughly 10$^{21}$~cm$^{-2}$,
which is several times higher than the typical HI column where our
sample of outer HII regions reside (a few times 10$^{20}$~cm$^{-2}$).}
This comparison suggests an intriguing  similarity in the mean
chemical enrichment level of DLA systems  and present-day outer
galactic disks, with DLA systems being only slightly more metal poor.

\begin{figure}
\plotone{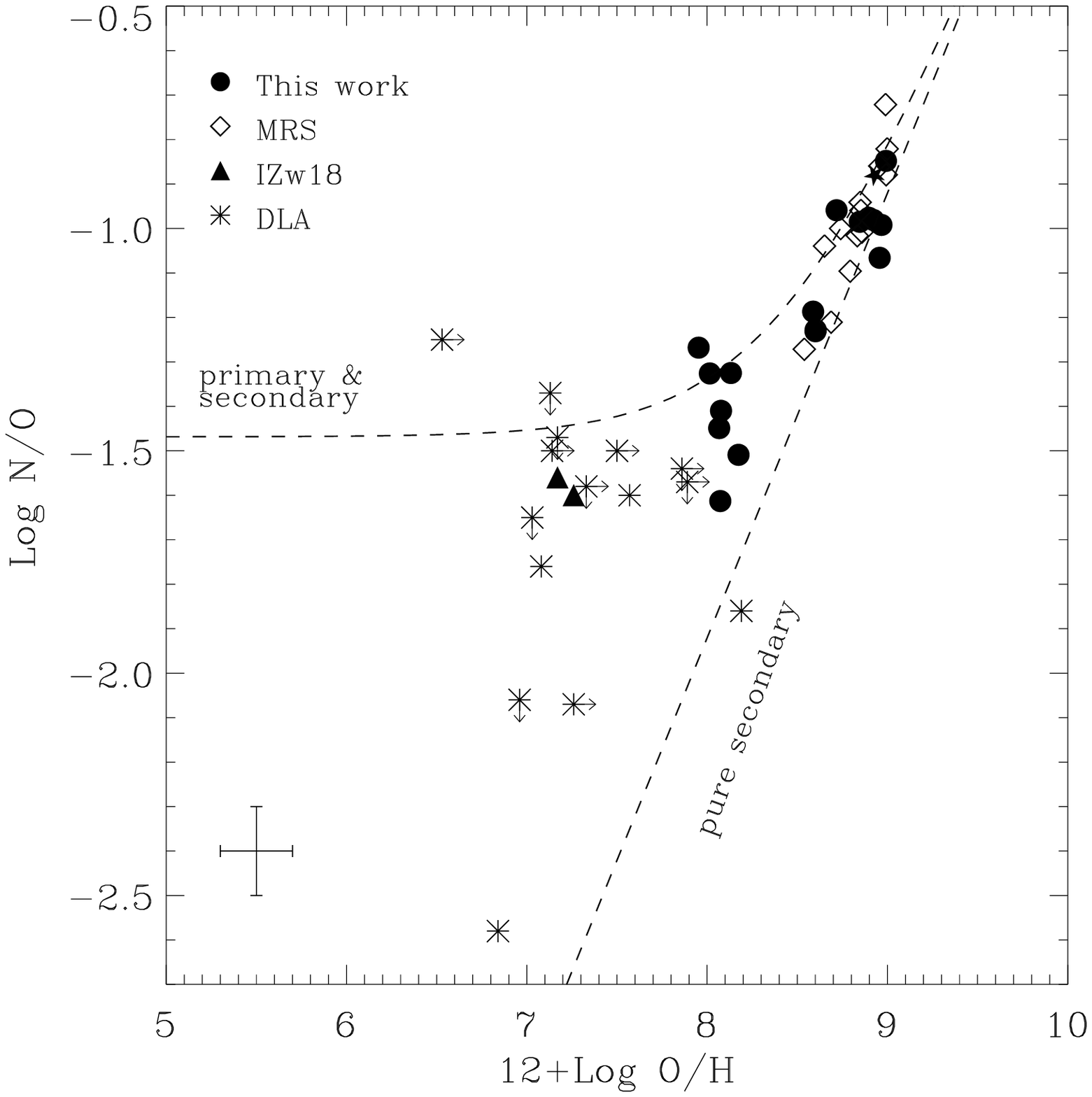}
\caption{Variation of log(N/O) vs 12+log(O/H) for our sample of HII
regions, as well as the inner disk HII regions measured by MRS.
Typical uncertainties in our measurements are indicated by the error
bar in the lower left.  The location of the Sun in this diagram is
indicated by a solid star.  We also plot values for the most metal poor
object known in the local Universe, IZw18 (Skillman \& Kennicutt 1993),
as well as the N/Si and Si/H measurements for a sample  of high
redshift DLA  systems from Lu et al (1998). (Note we have not shown the
error bars for these data, but only the upper and lower limits). The
dashed curves indicate the expectations for a pure secondary and
primary+secondary origins for nitrogen from Vila-Costas \& Edmunds
(1993).  The extreme outer disk HII regions are consistent with a
combination of both primary and secondary production of
nitrogen.\label{nooh}}
\end{figure}

Figure \ref{nooh} shows a a plot of  nitrogen-to-oxygen abundance as a
function of O/H  for our present sample of HII regions, as well as
those inner disk regions from MRS.  We have indicated where IZw18, the
most metal poor gas-rich object in the local Universe,  falls on this
diagram (Skillman \& Kennicutt 1993). Also shown are a set of  N/Si
measurements (or upper limits) for a sample of 15 DLA systems with
z$>$2  from Lu et al (1998). These authors argue that N/Si is
equivalent to N/O, under the assumption that O/Si is approximately
solar in DLA systems and that dust depletion is not significant.  This
plot reveals that while many DLA systems in the Lu et al sample are
considerably more metal poor than present-day outer galactic disks,
others  are only slightly displaced from outer disks in terms of O/H
and almost identical in terms of N/O.  The long dynamical timescales
($\sim$10$^8$--10$^9$~years) which characterize the extreme outer disks
imply that they evolve at a very slow rate but, nonetheless, we would
expect them to have been more metal poor in the past than at the
present epoch.  We therefore conclude that while, on average,   DLA
systems tend to be slightly more metal poor than outer galactic disks
at present, the two were very likely indistinguishable at large
lookback times.

\subsection{The Origin of Nitrogen}

A major unsolved question is  the importance of primary and secondary
processes in the production of nitrogen.  Nitrogen is believed to be
mostly a product of secondary nucleosynthesis, being produced via the
CNO cycle, however it is also thought to have primary component which
can be produced in the earlier helium burning stages (Renzini \& Voli
1981).  It is thought that secondary nitrogen is produced by stars of a
wide range of masses whereas primary nitrogen is produced by only
intermediate mass stars (4~$\lesssim$~M/M$_{\sun}$~$\lesssim$~8)
(Renzini \& Voli 1981, Matteucci 1986); there is evidence however that nitrogen
produced in massive stars is also predominantly primary. As an example,
Laird (1985) finds that [N/Fe] is constant for solar-neighbourhood
stars with -1.8$\le$[Fe/H]$\le$0.5, implying that nitrogen and iron are
produced in the same way, by the same stars.  Since it is well known
that some iron is primary, and produced on short timescales by Type II
supernovae from the core-collapse of massive stars (eg. Arnett 1995),
then one expects that such stars must also contribute to nitrogen
production.  The simple model for the chemical evolution of galaxies
(ie.  closed box) predicts that N/O will be independent of O/H for a
primary origin of nitrogen, and proportional to O/H for a secondary
origin (see Vila-Costas \& Edmunds 1993 for a discussion, including the
complication of $`$delayed' primary production).

The timescales for nitrogen production differ in stars of varying mass.
If nitrogen is predominantly produced by massive stars, then there
should be no time delay between the release of the element with respect
to oxygen,  and thus one expects the N/O ratio to exhibit only a small
amount of scatter. If, on the other hand, most nitrogen comes from
intermediate mass stars, then one expects a time delay of up to a few
times 10$^8$~years between the release of N/O. This is expected to
introduce a large scatter in the N/O ratio at low O/H, decreasing with
increasing metallicity as the effects of the time delay become less and
less important (eg. Garnett 1990; Pilyugin 1992).

In Figure \ref{nooh}, we have overplotted  the expectations for primary
and secondary production of nitrogen, as taken from Vila-Costas \&
Edmunds (1993).  At high metallicities, the expected linear trend
between O/H and N/O is seen, indicative of the dominant role of
secondary nitrogen production.  On the other hand, the extreme outer
disk HII regions are seen to  populate the region bracketed by the pure
secondary and primary + secondary curves, and thus appear consistent with
a combination of  primary and secondary production of nitrogen. The
spread in N/O at fixed O/H in the outer disk may very well reflect the
timescale between $`$bursts' of star formation, produced perhaps by the
passage of a spiral arm.   Further measurements, with improved
accuracy, are needed to better understand these results.

\section{Summary and Future Work}

We have presented the first results from a systematic study of the
physical properties and chemical abundances in a sample of
newly-discovered, extreme outer disk HII regions.  Optical spectra are
presented  for HII regions in  three, late-type spirals -- NGC~628,
NGC~1058 and NGC~6946 -- all of which are characterized by large than
average HI-to-optical sizes.   We have found that the outermost HII
regions studied, typically lying at 1.5--2\ropt,   have O/H abundances
in the range  of 9--15\% solar and  N/O abundances in the range of
20--25\% solar.  Evidence is also found for diminished dust extinction
at large radii, with the outermost HII regions having internal
extinctions of A$_V$~$\sim$~0-0.2.  Electron densities in the outer
disk HII regions are comparable to those found in the inner disk.  The
outer disk HII regions are observed to span a range in volume-averaged
ionization parameter $<$U$>$, and no correlation is seen with either
galactocentric radius or metallicity.

By combining our sample of outer disk measurements with those for inner
disk HII regions published in the literature, we have been able to
probe the radial variation of both oxygen and nitrogen-to-oxygen
abundances out to unprecedented radii.  Single log-linear relationships
are found to adequately describe the radial abundance variations,
although the derived slopes often differ considerably from those found
if only inner disk HII regions are used to define the fit.  The small
number of HII regions in our present sample, together with
uncertainties in the calibrations of the semi-empirical methods used
here to determine abundances,  limit the ability to constrain both
subtle changes in the radial gradient as well as scatter in the outer
disk.

Comparison of our outermost oxygen abundances with the predictions of
the simple closed-box model for chemical evolution reveals a general
consistency for two of the three galaxies, and suggests that the role
of gas flows in the evolution of their extreme outer disks is
comparable to that in the solar neighbourhood.  An intriguing
similarity is found between both the mean enrichment level and the
nitrogen-to-oxygen abundance in  outer disks and in some high redshift
DLA systems, implying that outer disks at the present epoch are
relatively unevolved systems.  While DLA systems tend to be slightly
more metal poor than present-day outer disks, they were very likely
indistinguishable at large lookback times.  Finally, we have found that
the outer disk HII regions in our sample are consistent with both
primary and secondary production of nitrogen.

A limitation of the current work is the reliance on semi-empirical
methods to determine the abundances. A concern is that the outer disk
HII regions under study here are physically different from the clusters
of OB stars on which the model calibrations are based. A more complete
set of photoionization models, extending to HII regions which are
ionized by  only a few massive stars, would be highly desirable.  We
are intending to measure the temperature-sensitive [OIII] $\lambda$4363
line, or at least place strong limits on it, for several of the
brightest HII regions in the present sample, in order to compare the
$`$direct' and semi-empirical abundance determinations for these
objects.  We are also continuing to obtain measurements of additional
HII regions in these galaxies, as well as extending our work to include
several other galaxies, residing in a range of physical environments
(eg. field and cluster).   The new generation of 8-m telescopes will
significantly ease the task of measuring  abundances for extreme outer
disk HII regions, while future HST instruments will make it possible to
routinely probe outer disk abundances via quasar absorption  lines.  A
larger sample of outer disk abundances will make it possible to study
the detailed nature of the abundance gradient at large radii, as well
as the amount of intrinsic scatter at a fixed radius.  Ultimately, we
will use our measurements of outer disk chemical abundances, along with
our measurements of past and present star formation rates, in order  to
construct self-consistent models of outer disk evolution.

\acknowledgements{We thank Stacy McGaugh for providing the grid
reproduced in Figure \ref{mg_grid} and for many useful discussions
concerning his model calibrations.   We thank Deidre Hunter, Tim
Heckman, Roberto Terlevich and Max Pettini for interesting discussions
and suggestions at various stages of this project.  AMNF acknowledges
support from an Amelia Earhart Fellowship from Zonta International.
JSG has been partially supported by JPL through the WFPC2 Investigation
Definition Team under NASA contract NAS 7-1260.  RFGW acknowledges
support from NASA grant NAGW-2892.}

\pagebreak


\begin{thebibliography}{dum}
\bibitem{} Afflerbach, A., Churchwell, E. \& Werner, M.~W. 1997, \apj, 478, 190
\bibitem{} Anders, N. \&  Grevesse,  E. 1989, \gca, 53, 197
\bibitem{} Arnett, D. 1995, \araa, 33, 115
\bibitem{} Belley, J. \& Roy, J.-R. 1992, \apjs, 78, 61
\bibitem{} Bingelli, B., Tamman, G.~A. \& Sandage, A. 1987, \aj, 94, 251\bibitem{} Briggs, F., Wolfe, A.~M., Liszt, H., Davis, M.~M. \& Turner, K.~L. 1989, \apj, 341, 650
\bibitem{} Broeils, A. 1994, Ph.~D. Thesis, Groningen University\bibitem{} Cayatte, V., Kotanyi, C., Balkowski, C. \& van Gorkam, J.~H. 1994, \aj, 107, 1003
\bibitem{} Clarke, C.~J. 1989, \mnras, 238, 283
\bibitem{} Czyak, S.~J., Keyes, C.~D. \& Aller, L.~H. 1986, \apjs, 61, 159
\bibitem{} De Geus, E.~J., Vogel, S.~N., Digel, S.~W. \& Gruendl, R.~A. 1993, \apj, 413, L97
\bibitem{} de Robertis, M.~M., Dufour, R.~J. \& Hunt, R.~W. 1987, JRASC, 85, 195
\bibitem{rc3}de Vaucouleurs, G., de Vaucouleurs, A., Corwin, H.~R., Buta, R.~J., Paturel, G., \&
Fouqu\`e, P.  1991, Third Reference Catalogue of Bright Galaxies (Springer-Verlag, New York) (RC3)
\bibitem{} Dickey, J.~M., Hanson, M. \& Helou, G. 1990, \apj, 352, 522 
\bibitem{} Djorgovski, S.~G., Pahre, M.~A., Bechtold, J. \& Elston, R. 1996, \nat, 382, 234
\bibitem{} Dopita, M.~A. \& Evans, I.~N. 1986, \apj, 307, 431
\bibitem{} Edmunds, M.~G. \& Pagel, B.~E.~J. 1984, \mnras, 211, 507
\bibitem{} Evans, I.~N. \& Dopita, M.~A. 1985, \apjs, 58, 125
\bibitem{} Ferguson, A.~M.~N., Wyse, R.~F.~G., Gallagher, J.~S. \& Hunter, D.~A. 1996a, \aj, 111, 2265
\bibitem{} Ferguson, A.~M.~N., Wyse, R.~F.~G. \& Gallagher, J.~S. 1996b, \aj, 112, 2567
\bibitem{} Ferguson, A.~M.~N. 1997, Ph.~D. Thesis, Johns Hopkins University.
\bibitem{} Ferguson, A.~M.~N., Wyse, R.~F.~G., Gallagher, J.~S. \& Hunter, D.~A. 1998a, ApJL, submitted.
\bibitem{} Ferguson, A.~M.~N., Wyse, R.~F.~G., Gallagher, J.~S. \& Hunter, D.~A 1998b, in preparation.
\bibitem{} Fich, M. \& Silkey, M. 1991, \apj, 366, 107
\bibitem{} Garnett, D.~R. 1990, \apj, 363, 142
\bibitem{} Garnett, D.~R. \& Shields, G.~A. 1987, \apj, 317, 82
\bibitem{} Garnett, D.~R., Odewahn, S.~C. \& Skillman, E.~D. 1992, \aj, 104, 1714
\bibitem{} Garnett, D.~R. \& Kennicutt, R.~C. 1994, \apj, 426, 123
\bibitem{} Garnett, D.~R., Shields, G.~A., Skillman, E.~D, Sagan, S.~P. \&
Dufour, R.~J. 1997a, \apj, 489, 63
\bibitem{} Garnett, D.~R., Skillman, E.~D,. Dufour, R.~J. \& Shields, G.~A. 1997b, \apj, 481, 174
\bibitem{} Gunn, J. \& Gott, J.~R. 1972, \apj, 176, 1
\bibitem{} Hodge, P. \& Kennicutt, R.~C. 1983, AJ, 88, 296
\bibitem{how83} Howarth, I. 1983, \mnras, 203, 301 
\bibitem{} Izotov, Y.~I., Thuan, T.~T. \& Lipovetsky, V.~A. 1994, \apj, 435, 647
\bibitem{} Kamphuis, J. 1993, Ph.~D. Thesis, Groningen University 
\bibitem{} Kauffmann, G., White, S.~D.~M. \& Guiderdoni, B. 1993, \mnras, 264, 201 
\bibitem{} Kennicutt, R.~C. \& Garnett, D.~R. 1996, \apj, 456, 504
\bibitem{} Kuijken, K. \& Gilmore, G. 1989, \mnras, 239, 605
\bibitem{} Laird, J.~B. 1985, \apj, 289, 556
\bibitem{} Larson, R.~B. 1976, \mnras. 176, 1
\bibitem{} Lu, L., Sargent, W.~L., Barlow, T.~A., Churchill, C.~W. \& Vogt,
S.~S. 1996, \apjs, 107, 475
\bibitem{} Lu, L., Sargent, W.~L. \& Barlow, T.~A 1998, \aj, 115, 55
\bibitem{} Massey, P., Strobel, K., Barnes, J.~V. \& Anderson, E. 1988, \apj, 328, 315
\bibitem{} Matteucci, F. 1986, \mnras, 211, 911
\bibitem{} McCall, M.~L., Rybski, P.~M. \& Shields, G.~A. 1984, \apjs, 57, 1 (MRS)
\bibitem{} McGaugh, S.~S. 1991, \apj, 380, 140
\bibitem{} McGaugh, S.~S. 1994, \apj, 426, 135
\bibitem{} Mendoza, C. \& Zeippen, C.~J. 1982, \mnras, 198, 127
\bibitem{} Miller, B.~W. 1994, Ph.~D. Thesis, University of Washington
\bibitem{} Miller, B.~W. \& Hodge, P. 1996, \apj, 458, 467
\bibitem{} Molla, M., Ferrini, F. \& Diaz, A.~I. 1996, \apj, 466, 668
\bibitem{} Nussbaumer, H. \& Storey, P.~J. 1981, \aap, 96, 91
\bibitem{} Oey, M.~S. \& Kennicutt, R.~C. 1993, \apj, 411, 1370
\bibitem{ost89}Osterbrock, D.~E. 1989, Astrophysics of Gaseous Nebulae and Active Galactic Nuclei
(Mill Valley: University Science Books)
\bibitem{} Pagel, B.~E.~J., Edmunds, M.~G., Blackwell, D.~E., Chun, M.~S. \& Smith, G. 1979, \mnras, 189, 95
\bibitem{} Pagel, B.~E.~J., Edmunds, M.~G. \& Smith, G. 1980, \mnras, 193, 219
\bibitem{}Pagel, B.~E.~G. \& Edmunds, M.~G. 1981, \araa, 2, 77
\bibitem{} Pettini, M., Smith, L.~J., Hunstead, R.~W. \& King, D.~L. 1994, \apj, 426, 79
\bibitem{} Pettini, M., Lipman, K., \& Hunstead, R.~W. 1995, \apj, 451, 100
\bibitem{} Pettini, M., Smith, L.~J., King, D.~L. \& Hunstead, R.~W. 1997,
\apj, 486, 665
\bibitem{} Pickering, T., Impey, C., van Gorkom, J. \& Bothun, G. 197, \aj, 114, 1858
\bibitem{} Pilyugin, L. 1992, \aap, 260, 58
\bibitem{} Prantzos, N. \& Aubert, O. 1995, \aap, 302, 69
\bibitem{} Renzini, A. \& Voli, M. 1981, \aap, 94, 175
\bibitem{} Rudolph, A.~L., Simpson, J.~P., Haas, M.~R., Erickson, E.~F. \& Fich, M. 1997, \apj, 489, 94
\bibitem{} Ruphy, S., Robin, A.~C., Epchtein, N., Copet, E., Bertin, E., Fouque, P. \& Gugliemo, F. 1996, \aap, 313, 21
\bibitem{}  Schechter, P.~L. 1980, \aj, 85, 801
\bibitem{schi77} Schild, R.~E. 1977, \aj, 82, 337
\bibitem{} Schmidt, M. 1963, \apj, 137, 758
\bibitem{sco92} Scowen, P.~A., Dufour, R.~J. \& Hester, J.~J. 1992, \aj, 104, 92
\bibitem{} Searle, L. 1971, \apj, 168, 327
\bibitem{} Searle, L. \& Sargent, W.~L.~W. 1972, \apj, 173, 25
\bibitem{} Seaton, M.~J. 1979, \mnras, 185, 57
\bibitem{} Shields, G.~A. 1974, \apj, 193, 335
\bibitem{} Shields, J.~C. \& Kennicutt, R.~C. 1995, \apj, 454, 807
\bibitem{} Shostak, G.~S. \& van der Kruit, P.~C. 1984, \aap, 132, 20
\bibitem{} Skillman, E.~D. 1989, \apj, 347, 883
\bibitem{} Skillman, E.~D., Kennicutt, R.~C. \& Hodge, P.~W. 1989, \apj, 347, 875
\bibitem{} Skillman, E.~D., Terlevich, R.~J., Kennicutt, R.~C., Garnett, D.~R. 
\& Terlevich, E. 1994, \apj, 431, 172
\bibitem{} Skillman, E.~D. \& Kennicutt, R.~C. 1993, \apj, 411, 655
\bibitem{} Skillman, E.~D., Bomans, D.~J. \& Kobulnicky, H.~A. 1997, \apj,
474, 205
\bibitem{} Thuan, T., Izotov, Y.~I. \& Lipovetsky, V.~A. 1995, \apj, 445, 108 
\bibitem{} Thurston, T.~R., Edmunds, M.~G. \& Henry, R.~B.~C. 1996, \mnras, 283, 990
\bibitem{vgs96} Vacca, W.~D., Garmany, C.~D. \& Shull, J.~M. 1996, \apj, 460, 914
\bibitem{} van der Kruit, P.~C. \& Shostak, G.~S. 1984, \aap, 134, 258
\bibitem{} Vila-Costas, M.~B. \& Edmunds, M.~G. 1992, \mnras, 259, 121
\bibitem{} Vila-Costas, M.~B. \& Edmunds, M.~G. 1993, \mnras, 265, 199
\bibitem{} Vilchez, J.~M., Pagel, B.~E.~J., Diaz, A.~I., Terlevich, E.
\& Edmunds, M.~G. 1988, \mnras, 235, 633
\bibitem{} Vilchez, J.~M. \& Esteban, C. 1996, \mnras. 280, 720 
\bibitem{ws83}Webster, B.~L. \& Smith, M.~G. 1983, \mnras, 204, 743
\bibitem{} White, S.~D.~M. \& Rees, M.~J. 1987, \mnras, 183, 341
\bibitem{} Wolfe, A. 1988 in Quasar Absorption Lines, eds. J.~C. Blades, D. Turnshek \& C.~A. Norman (Cambridge University Press), p. 297
\bibitem{} Wolfe, A.~M., Lanzetta, K.~M., Foltz, C.~B. \& Chaffee, F.~H. 1995,
\apj, 454, 698
\bibitem{} Wyse, R.~F.~G. \& Silk, J. 1989, \apj, 339, 700
\bibitem{} Wyse, R.~F.~G. \& Gilmore, G. 1995, \aj, 110, 2771
\bibitem{zkh94} Zaritsky, D., Kennicutt, R.~C., \& Huchra, J.~P. 1994, \apj, 420, 558 (ZKH)
\bibitem{} Zaritsky, D. 1995, \apj, 448, L17

\end{thebibliography}
\end{document}